\documentclass[graybox]{svmult}

\usepackage{type1cm}        
\usepackage{makeidx}         
\usepackage{graphicx}        
\usepackage{multicol}        
\usepackage[bottom]{footmisc}

\usepackage{newtxtext}       %
\usepackage[varvw]{newtxmath}       

\makeindex             

\usepackage{nicefrac}
\usepackage[T1]{fontenc}
\usepackage[utf8]{inputenc}
\usepackage[dvipsnames]{xcolor}
\usepackage[numbers,sort&compress]{natbib}
\usepackage[
    colorlinks=true,
    linkcolor=ForestGreen,
    citecolor=MidnightBlue,
    urlcolor=BrickRed
]{hyperref}
\usepackage{doi}

\usepackage{appendix}

\begin{document}

\newcommand{\ER}{Erd\H{o}s-R\'enyi}

\title*{A Guide to Higher-Order Homophily}


\author{Moritz Laber\orcidID{0000-0002-0273-584X} and\\ Brennan Klein\orcidID{0000-0001-8326-5044}}
\institute{Moritz Laber \at Northeastern University, 360 Huntington Ave, Boston, USA, \email{laber.m@northeastern.edu}
\and Brennan Klein \at Northeastern University, 360 Huntington Ave, Boston, USA, \email{b.klein@northeastern.edu}}


\maketitle

\abstract*{Homophily, the overrepresentation of interactions among similar individuals, and heterophily, the elevated prevalence of interactions among dissimilar ones, are frequently observed mixing patterns in social networks. As hypergraphs are increasingly used to represent social systems, a higher-order perspective on homophily and heterophily becomes ever more relevant. Here, we provide two complementary perspectives on this problem: First, we survey measures that can be used to quantify homophily (or heterophily) in hypergraphs---emphasizing conceptual differences to existing pairwise measures---and explain each measure through in-depth examples. Second, we provide an overview of hypergraph models for higher-order mixing patterns, distinguishing several model families with distinct use cases. By providing a guide to existing methods and synthesizing the current body of knowledge on higher-order homophily and heterophily, we lay the basis for informed methodological choices and future developments.}

\abstract{Homophily, the overrepresentation of interactions among similar individuals, and heterophily, the elevated prevalence of interactions among dissimilar ones, are frequently observed mixing patterns in social networks. As hypergraphs are increasingly used to represent social systems, a higher-order perspective on homophily and heterophily becomes ever more relevant. Here, we provide two complementary perspectives on this problem: First, we survey measures that can be used to quantify homophily (or heterophily) in hypergraphs---emphasizing conceptual differences to existing pairwise measures---and explain each measure through in-depth examples. Second, we provide an overview of hypergraph models for higher-order mixing patterns, distinguishing several model families with distinct use cases. By providing a guide to existing methods and synthesizing the current body of knowledge on higher-order homophily and heterophily, we lay the basis for informed methodological choices and future developments.}

\section{Introduction}

Homophily, the above-chance occurrence of interactions between similar individuals, and its counterpart, heterophily, the above-chance occurrence of interactions among dissimilar individuals, are core dimensions along which to examine the structure of social networks~\cite{mcpherson2001_birdsfeather, bojanowski2014_measuringsegregation, lawrence2020_homophilymeasuresmeaning}. The study of such \textit{mixing patterns}~\cite{newman2003_mixingpatterns}---the term we use to refer to the dependence of interactions on the attributes of the individuals participating in them, irrespective of whether they are homophilous or heterophilous---has been the focus of a large body of scholarly work dating back to the first half of the twentieth century~\cite{mcpherson2001_birdsfeather, freeman1996_antecedentssocialnetwork}. Mixing patterns have been studied with respect to a myriad of attributes, including political ideology, professional specialization, education, race, gender, and many more (see, e.g.,~\cite{mcpherson2001_birdsfeather} for an overview of empirical results). However, distinguishing whether an observed mixing pattern is due to individual preferences or arises from unequal attribute prevalence or alternative mechanisms has been recognized as a key challenge~\cite{lawrence2020_homophilymeasuresmeaning, rivera2010_dynamicsdyads, peixoto2022_disentanglinghomophily}.

While the existence of nontrivial mixing patterns in social networks is seldom challenged, questions about their meaning~\cite{lawrence2020_homophilymeasuresmeaning}, their impact on network formation~\cite{rivera2010_dynamicsdyads}, and their consequences for individual outcomes~\cite{ertug2022_whatdoeshomophilydo} are still topics of scholarly debate.

The sustained interest in analyzing social systems through the lens of higher-order networks~\cite{battiston2025_collectivehumanbehavior, stonge2025_modelsofgroups} naturally calls for a higher-order perspective on mixing patterns. However, the definition, measurement, and modeling of homophily and heterophily in the context of higher-order networks all pose unique challenges above and beyond those present in networks of pairwise interactions. For example, consider a network where nodes can be in one of two classes: purple or green. Each edge in this network connects either two nodes of the same color or two nodes of different colors. This is no longer true in hypergraphs, as hyperedges can contain an arbitrary number of nodes of either color. Not only does this invalidate the dichotomy of ``within-group'' and ``between-group'' edges (see e.g.,~\cite{kaminski2024_modularity-2, veldt2023_hypergraphhomophily}), but it also leads to combinatorial constraints on the mixing patterns one could possibly observe in a hypergraph~\cite{veldt2023_hypergraphhomophily}. These constraints, together with the combinatorial explosion of possible interactions and graphicality conditions that jointly constrain node degrees and hyperedge sizes, make the design of statistical and computational models of hypergraphs with mixing patterns a unique challenge. In addition to these fundamental concerns, there is a practical one: While extensive reviews offer synthesis and guidance in the case of pairwise networks (e.g., \cite{mcpherson2001_birdsfeather, bojanowski2014_measuringsegregation, lawrence2020_homophilymeasuresmeaning}), there are few similarly detailed reviews on homophily and heterophily in higher-order networks~\cite{saxena2025_homophilytutorial}.

The goal of this chapter is to serve as a guide to mixing patterns in higher-order networks by first surveying measures for higher-order homophily and heterophily (Section~\ref{sec:homophily-measures}) and, second, by providing an overview of hypergraph models for the statistical and computational investigation of higher-order mixing patterns (Section~\ref{sec:hypergraph-models}). 

When reviewing measures of higher-order mixing patterns, we take the perspective that quantifying homophily (or heterophily) means measuring the above-chance occurrence of interactions between similar (or dissimilar) individuals~\cite{lawrence2020_homophilymeasuresmeaning}. Throughout, we emphasize that it is necessary to specify (1) what it means for two or more individuals to interact, (2) what it means for them to be similar, and (3) with respect to which null model ``above-chance'' is to be understood. For each reviewed measure, we specify its definition, provide an illustrative example, and discuss the conceptual insights it offers as well as its limitations.

Our introduction to hypergraph models aims to expose the reader to the main classes of models that we believe are relevant to the contemporary study of higher-order mixing patterns. This includes null models, statistical models of hypergraphs with discrete or continuous node attributes, and the broad category of mechanistic and growing models of hypergraphs with controllable mixing patterns. We distinguish these classes because they serve different purposes: Null models specify a baseline against which the structure of empirical data can be compared~\cite{cimini2019_statisticalphysics, preti2024_higherordernullmodels}. Statistical models specify a (simple) probabilistic generative mechanism in terms of a few interpretable parameters that can be learned from data to gain insight into a social system of interest and may permit sampling of new hypergraphs given the parameters~\cite{crane2018_probabilisticfoundations, peel2022_statisticalinference}. Finally, mechanistic and growing hypergraph models specify a generative process in terms of algorithmic rules, often with the goal of exploring the impact of a particular hyperedge formation rule or the desire to efficiently create hypergraphs that resemble realistic properties found in hypergraph data (see, e.g.,~\cite{giroire2022_PAhypergraphs, laber2025_effectshigherorderinteractions}).

We believe that designing measures of higher-order mixing patterns and the modeling of hypergraph structure are two fundamental problems that research into higher-order homophily and heterophily has to address, as they form the foundation of robust empirical studies and our understanding of the consequences of higher-order homophily for individual outcomes in social systems.


\section{Quantifying Higher-Order Mixing Patterns}\label{sec:homophily-measures}

Here, we survey approaches for quantifying the extent to which interactions in a higher-order network are homophilous or heterophilous. In the remainder of this section, we introduce key notation needed to define these measures. The following sections each introduce a different measure, highlighting its key conceptual contributions as well as potential limitations. 

Specifying a \textit{homophily measure} requires defining interactions, similarity, and a null model. We use the term \textit{homophily} as a property of interaction structure, following many recent papers on higher-order social systems, but emphasize that this term has historically been used with a more agentic connotation, emphasizing the preference of individuals for interactions with similar individuals. The homophily measures we review, thus, quantify mixing patterns in a broad sense, and we include measures like assortativity and modularity. We also note that most measures can assess, despite their name, the presence (or absence) of homophily as well as heterophily.

In this chapter, \textit{interactions} among individuals are defined through hyperedges $e \in \mathcal{E}$ among nodes $i\in \mathcal{V}$ in a hypergraph $\mathcal{H}=(\mathcal{V},\mathcal{E})$. Admittedly, this defers the hard problem of deciding what type of real-world relationship constitutes a hyperedge. We use $\mathcal{E}_\alpha$ to denote the set of hyperedges of size $\alpha$, and call two hyperedges that contain the same nodes \textit{parallel}. In this case, $\mathcal{E}$ is a multiset. Following common nomenclature, we call a hyperedge \textit{simple} if each node appears at most once in it. A hypergraph is \textit{simple} if it contains only simple and non-parallel hyperedges.

We assume that the \textit{similarity} of nodes is a function of their attributes. We will encounter two types of attributes: categorical (more precisely, nominal) attributes and continuous attributes. For categorical attributes, we focus on the case in which each node $i \in \mathcal{V}$ is assigned one of $C$ attributes $c_i$ from an unordered set $\mathcal{C}$. We think of these attributes as the labels of different classes to which nodes may belong and denote the set of nodes in class $c$ as $\mathcal{V}_c$. In the context of pairwise networks, these classes are often called ``groups'' or ``communities''. We avoid this terminology as group is easily confused with group interaction, i.e., hyperedge, and community often presupposes that its members connect homophilously, which we do not assume a priori. In the literature, hyperedges that contain only members from one class are sometimes called ``pure,'' and hyperedges that contain members from different classes are referred to as ``diverse''. As both terms are value-laden, we avoid them and prefer the more neutral terms \textit{homogeneous} and \textit{heterogeneous}, respectively. For continuous and potentially multi-dimensional attributes, we denote the attribute of node $i$ as $x_i \in \mathcal{X}$ and will specify $\mathcal{X}$ in the relevant context. It is sometimes useful to think of such attributes as node positions in a latent space. 
In the case of discrete attributes, all homophily measures we review regard nodes as similar if they belong to the same class. In the case of continuous attributes, distances or inner products are common measures of similarity.

Measuring homophily, according to our definition, requires a \textit{null model} to determine whether interactions between similar individuals are more present than one would expect ``by chance''. The homophily measures we review often define a null model explicitly but some incorporate it implicitly (e.g., through the use of correlation coefficients) or even omit it, in which case the measure does not strictly fall under our definition but could still be used to compare different systems relative to each other.

\subsection{Hypergraph Homophily}\label{sec:hypergraph-homophily}

In their foundational work, Veldt, Benson, and Kleinberg (VBK)~\cite{veldt2023_hypergraphhomophily} define a family of measures for homophily in hypergraphs as a generalization of the homophily index for pairwise interactions~\cite{coleman1958_homophily, currarini2009_economicmodelfriendship, altenburger2018_monophily}. They show that, in contrast to the pairwise case, not all possible values of this measure can be realized in hypergraphs due to combinatorial constraints.

Here, we first introduce VBK's family of hypergraph homophily measures, followed by a worked example, and finally highlight important practical considerations, including the combinatorial impossibility result.

\subsubsection{Definition}

Hyperedges of size $\alpha > 2$ do not follow the ``within-class'' and ``between-class'' dichotomy. They are instead characterized by their class composition, i.e., how many nodes of a given class they contain. The concept of \textit{type} $t_c(e)$ of a hyperedge $e$ relative to class $c$, defined as the number of nodes of class $c$ that are members of $e$, formalizes this idea,
\begin{equation}\label{eq:type}
    t_c(e)
    =
    \left| \{ i \in e: c_i = c \} \right|
    .
\end{equation}
We refer to a hyperedge $e$ of size $\alpha$ and type $t_c(e) = t$ relative to class $c$ as a $(c, \alpha, t)$-hyperedge, and denote the number of such hyperedges in the hypergraph as $E^{(c)}_{\alpha, t}$. Similarly, we define the $(c, \alpha, t)$-degree of node $i$ relative to $c$ as the number of $(c, \alpha, t)$-hyperedges node $i$ participates in, i.e.,
\begin{equation}\label{eq:degree-size-type-dependent}
    d^{(c)}_{\alpha, t}(i)
    =\sum_{e \in \mathcal{E}} \mathbf{1}\{ i \in e ~ \wedge ~ |e| = \alpha ~ \wedge ~ t_c(e) = t \},
\end{equation}
where $\mathbf{1} \{\, \cdot \,\}$ denotes the indicator function.

With these definitions in place, we can turn to the central quantities of hypergraph homophily, the \textit{standard type-$t$ affinity scores} $a_{c, \alpha, t}^{\mathrm{VBK}1}$ and the \textit{alternative type-$t$ affinity scores} $a_{c, \alpha, t}^{\mathrm{VBK}2}$ for hyperedge size $\alpha$ and with respect to reference class $c$. The former captures the contribution of $(c, \alpha, t)$-hyperedges to the total number of hyperedges of size $\alpha$ that nodes in class $c$ participate in,
\begin{equation}\label{eq:vbk-standard-affinity-scores}
    a_{c, \alpha, t}^{\mathrm{VBK}1}
    =
    \frac{
            \sum_{i \in \mathcal{V}_c} d^{(c)}_{\alpha, t}(i)
        }
        {
            \sum_{i \in \mathcal{V}_c} d_{\alpha}(i)
        }
    =
    \frac{
        t E^{(c)}_{\alpha, t}
    }
    {
        \sum_{t' = 1}^{\alpha} t' E^{(c)}_{\alpha, t'}
    },
\end{equation}
where $d_\alpha(i)$ is the degree of node $i$ only taking into account size $\alpha$ hyperedges, i.e.,
\begin{equation}\label{eq:degree-size-dependent}
    d_{\alpha}(i)
    =
    \sum_{e \in \mathcal{E}} \mathbf{1} \{ i \in e ~\wedge~ |e| = \alpha \}
    = 
    \sum_{t=0}^{\alpha} d^{(c)}_{\alpha, t}(i),
\end{equation}
which is independent of the reference class $c$. The factor $t$ in Eq.~\eqref{eq:vbk-standard-affinity-scores} accounts for the fact that a hyperedge $e$ of type $t_c(e) = t$ contributes to the degree of $t$ nodes in class $c$.

The alternative affinity scores $a_{c, \alpha, t}^{\mathrm{VBK}2}$ disregard this factor $t$ and simply measure affinity in terms of the relative abundance of $(c, \alpha, t)$-hyperedges in the hypergraph, 
\begin{equation}\label{eq:vbk-alternative-affinity-scores}
    a_{c, \alpha, t}^{\mathrm{VBK}2}
    =
    \frac{
            E^{(c)}_{\alpha, t}
        }
        {
            \sum_{t' = 1}^{\alpha} E^{(c)}_{\alpha, t'}
        }.
\end{equation}
Note, however, that the alternative scores are still class dependent due to the denominator, which only takes into account hyperedges with at least one member of the reference class. This also means that both standard and alternative affinity scores are only defined for $t \geq 1$.

To judge whether the interactions of size $\alpha$ in a given hypergraph exhibit homophily or heterophily the affinity scores are compared to their expected value under an appropriate null model. In the case of the standard scores, the null model consists of choosing a node from the reference class $c$ and sampling the remaining $\alpha - 1$ nodes uniformly at random from all remaining $N - 1$ nodes. The \textit{standard type-$t$ baseline scores} at hyperedge size $\alpha$ and with respect to class $c$ are the probabilities of creating a $(c, \alpha, t)$-hyperedge with this procedure, 
\begin{equation}\label{eq:vbk-standard-baseline-scores}
    b_{c, \alpha, t}^{\mathrm{VBK}1}
    =
    \frac{
        \binom{N_c - 1}{t - 1}
        \binom{N - N_c}{\alpha - t}
    }
    {
        \binom{N - 1}{\alpha - 1}
    }
    \overset{
        N \to \infty
    }{
        \longrightarrow
    }
    \binom{\alpha - 1}{t - 1} \nu_c^{t - 1} (1 - \nu_c)^{\alpha - t}
    =
    \mathrm{Binom}[\alpha - 1, \nu_c](t - 1),
\end{equation}
where by assumption the fraction of class-$c$ nodes $\nu_c = \frac{N_c}{N}$ stays constant as $N \to \infty$ and $\mathrm{Binom}[\alpha - 1, \nu_c](t - 1)$ is the probability mass function of the Binomial distribution, i.e., the probability of obtaining $t - 1$ successes when sampling $\alpha - 1$ Bernoulli trials with success probability $\nu_c$.

For the alternative scores, the null model consists of sampling hyperedges uniformly at random from all size-$\alpha$ hyperedges with at least one node from the reference class $c$. The \textit{alternative type-$t$ baseline scores} at hyperedge size $\alpha$ and with respect to class $c$ are the probabilities of obtaining a $(c, \alpha, t)$-hyperedge through this procedure, 
\begin{equation}\label{eq:vbk-alternative-baseline-scores}
    b_{c, \alpha, t}^{\mathrm{VBK}2}
    = 
    \frac{
            \binom{N_c}{t}
            \binom{N - N_c}{\alpha - t}
        }{
            \sum_{t' = 1}^{\alpha} \binom{N_c}{t'}\binom{N - N_c}{\alpha - t'}
        }
    \overset{
            N \to \infty
        }{
            \longrightarrow
        }
    \frac{
            \binom{\alpha}{t} \nu_c^t (1 - \nu_c)^{\alpha  - t}
        }{
            1 - (1 - \nu_c)^\alpha
        }
    =
    \frac{
            \mathrm{Binom}[\alpha, \nu_c](t)
        }{
            \sum_{t'=1}^{\alpha}\mathrm{Binom}[\alpha, \nu_c](t')
        },
\end{equation}
where by assumption $\nu_c = \frac{N_c}{N}$ stays constant as $N \to \infty$ and the denominator of the asymptotic form results from conditioning the Binomial distribution on at least one success. For both standard and alternative scores the asymptotic baseline scores offer greater numerical stability than the exact expression when analyzing large hypergraphs.

Different strategies can be used to compare affinity scores to baseline scores: The \textit{ratio scores} divide the affinity score by the baseline score, while the \textit{normalized bias scores} subtract the baseline score from the affinity score and divide by the maximum possible difference. This gives rise to four different hypergraph homophily measures: The \textit{standard ratio scores}
\begin{equation}\label{eq:vbk-standard-ratio-score}
    h_{c, \alpha, t}^{\mathrm{VBK}1}
    =
    \frac{
        a_{c, \alpha, t}^{\mathrm{VBK}1}
    }{
        b_{c, \alpha, t}^{\mathrm{VBK}1}
    }
    , 
\end{equation}
the \textit{alternative ratio scores} 
\begin{equation}\label{eq:vbk-alternative-ratio-score}
    h_{c, \alpha, t}^{\mathrm{VBK}2}
    =
    \frac{
        a_{c, \alpha, t}^{\mathrm{VBK}2}
    }{
        b_{c, \alpha, t}^{\mathrm{VBK}2}
    }
    ,
\end{equation}
the \textit{standard normalized bias scores}
\begin{equation}\label{eq:vbk-standard-bias-score}
    h_{c, \alpha, t}^{\mathrm{VBK}1\mathrm{B}}
    =
    \begin{cases}
        & \frac{
            a_{c, \alpha, t}^{\mathrm{VBK}1} - b_{c, \alpha, t}^{\mathrm{VBK}1}
        }{
            1 - b_{c, \alpha, t}^{\mathrm{VBK}1}}
        ~~\text{if}~~ a_{c, \alpha, t}^{\mathrm{VBK}1} \geq b_{c, \alpha, t}^{\mathrm{VBK}1} \\
        & \frac{
                a_{c, \alpha, t}^{\mathrm{VBK}1} - b_{c, \alpha, t}^{\mathrm{VBK}1}
            }{
                b_{c, \alpha, t}^{\mathrm{VBK}1}
            }
        ~~\text{if}~~ a_{c, \alpha, t}^{\mathrm{VBK}1} < b_{c, \alpha, t}^{\mathrm{VBK}1},
    \end{cases}
\end{equation}
and finally, the \textit{alternative normalized bias scores}
\begin{equation}\label{eq:vbk-alternative-bias-score}
    h_{c, \alpha, t}^{\mathrm{VBK}2\mathrm{B}}
    =
    \begin{cases}
        & \frac{
            a_{c, \alpha, t}^{\mathrm{VBK}2} - b_{c, \alpha, t}^{\mathrm{VBK}2}
        }{
            1 - b_{c, \alpha, t}^{\mathrm{VBK}2}}
        ~~\text{if}~~ a_{c, \alpha, t}^{\mathrm{VBK}2} \geq b_{c, \alpha, t}^{\mathrm{VBK}2} \\
        & \frac{
                a_{c, \alpha, t}^{\mathrm{VBK}2} - b_{c, \alpha, t}^{\mathrm{VBK}2}
            }{
                b_{c, \alpha, t}^{\mathrm{VBK}2}
            }
        ~~\text{if}~~ a_{c, \alpha, t}^{\mathrm{VBK}2} < b_{c, \alpha, t}^{\mathrm{VBK}2}.
    \end{cases}
\end{equation}

The authors develop a classification scheme for hypergraphs, mirroring existing graph-based schemes~\cite{currarini2009_economicmodelfriendship}, based on the standard ratio scores. However, it is easy to imagine similar schemes devised based on the three other scores.

A given class $c$ exhibits \textit{simple homophily} at hyperedge size $\alpha$ if
\begin{equation}\label{eq:vbk-simple-homophily}
    h_\mathrm{c,\alpha, \alpha}^{\mathrm{VBK}1}
    >
    1
    ,
\end{equation}
i.e., homogeneous hyperedges are overrepresented with respect to the standard baseline score. While this definition is attractive due to its simplicity, it does not account for the structural information in heterogeneous hyperedges. Furthermore, creating homogeneous hyperedges becomes increasingly unlikely as hyperedge size increases, both under the null model and, conceivably, also in reality.

Extending this idea to mixed hyperedges yields the concept of \textit{majority homophily}. Class $c$ exhibits order-$j$ majority homophily at hyperedge size $\alpha$ if
\begin{equation}\label{eq:vbk-majority-homophily}
    h_{c, \alpha, \alpha - j' + 1}^{\mathrm{VBK}1}
    >
    1
    ~~ \text{for} ~~
    j' 
    \leq
    j
    .
\end{equation}
The largest $j$ for which this condition is satisfied is called the \textit{majority homophily index} (MaHI). If the MaHI is $j = \lceil \frac{\alpha}{2} \rceil$, class $c$ exhibits \textit{strict majority homophily} at hyperedge size $\alpha$.

A different notion of homophily is satisfied if the standard ratio scores are monotonically increasing,
\begin{equation}
    h_{c, \alpha, \alpha - j +1}^{\mathrm{VBK}1}
    < 
    h_{c, \alpha, \alpha - (j -1) +1}^{\mathrm{VBK}1}
    <
    \dots
    <
    h_{c, \alpha, \alpha}^{\mathrm{VBK}1}.
\end{equation}
This condition defines \textit{order}-$j$ \textit{monotonic homophily} for class $c$ at hyperedge size $\alpha$. The largest $j$ such that monotonic homophily is satisfied is called the \textit{monotonic homophily index} (MoHI). If the MoHI is $j=\lceil \frac{\alpha}{2} \rceil$, one says that class $c$ exhibits \textit{strict monotonic homophily} at hyperedge size $\alpha$.

\subsubsection{Example}

\begin{figure}[tb]
    \centering
    \includegraphics[width=0.99\linewidth]{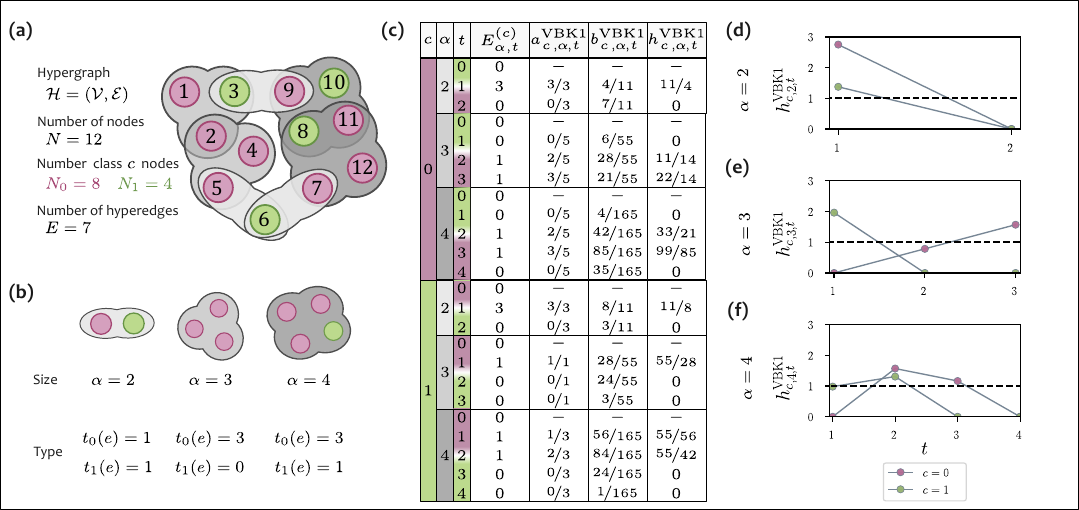}
    \caption{\textbf{Hypergraph homophily:}
    \textbf{(a)} The hypergraph $\mathcal{H}$ has $N=12$ nodes, $N_0=8$ of which are in class $0$ (purple) and $N_1=4$ are in class $1$ (green). The total number of hyperedges is $E=7$.
    \textbf{(b)} The type $t_c(e)$ of different hyperedges $e$ with respect to class $c \in \{0,1\}$ is the number of class $c$ nodes in them.
    \textbf{(c)} The standard ratio scores $h_{c, \alpha, t}^{\mathrm{VBK}1}$ for $(c, \alpha, t)$-hyperedges are calculated from the number of such hyperedges $E^{(c)}_{\alpha, t}$ for fixed size $\alpha$ and reference class $c$ but varying type $t \in \{1, \dots, \alpha \}$ as well as the class sizes $N_c$. Inspecting the standard ratio scores $h_{c, \alpha, t}^{\mathrm{VBK}1}$ for $c=0$ (purple circles) and $c=1$ (green circles) as a function of the type $t$ in comparison to the neutral baseline $h_{c, \alpha, t}^{\mathrm{VBK}1}=1$ (dashed black line) shows \textbf{(d)} heterophily for both classes at hyperedge size $\alpha = 2$, \textbf{(e)} strict monotonic homophily for class $c=0$ but not class $c=1$ at hyperedge size $\alpha = 3$, and \textbf{(f)} heterophily at hyperedge size $\alpha = 4$.}
    \label{fig:example-hypergraph}
\end{figure}

To illustrate the above definitions, we apply them to the hypergraph in Fig.~\ref{fig:example-hypergraph}\textbf{(a)}. This hypergraph has $N=12$ nodes labeled by positive integers and $E = 7$ hyperedges of different sizes, $\alpha \in \{2,3,4\}$. The nodes belong to one of two classes $\mathcal{C} = \{ 0, 1 \}$. We observe class imbalance, because $N_0=8$ nodes belong to class $0$ (purple), while only $N_1 = 4$ nodes belong to class $1$ (green). 

The size $\alpha$ of the example hyperedges in Fig.~\ref{fig:example-hypergraph}\textbf{(b)} is determined by the total number of nodes in them. The type $t_0(e)$ of a hyperedge $e$ with respect to class $0$ is the number of class $0$ (purple) nodes in $e$, and the type $t_1(e)$ with respect to class $1$ is the number of class $1$ (green) nodes in $e$. The example  hyperedges illustrate that the dichotomy between within-class and between-class edges breaks down in hypergraphs, and needs to be replaced with the more flexible notion of type.

Figure~\ref{fig:example-hypergraph}\textbf{(c)} shows how to compute the standard ratio scores (Eq.~\eqref{eq:vbk-standard-ratio-score}). The table contains the number of hyperedges $E^{(c)}_{\alpha, t}$ of size $\alpha$ and type $t$ with respect to class $c$ for all possible sizes, types, and reference classes. The table is not fully populated as some combinations of type and size are not observed. The hyperedge counts provide all information needed to compute the standard affinity scores $a_{c, \alpha, t}^{\mathrm{VBK}1}$ (Eq.~\eqref{eq:vbk-standard-affinity-scores}). Computing the standard baseline scores $b_{c, \alpha, t}^{\mathrm{VBK}1}$ (Eq.~\eqref{eq:vbk-standard-baseline-scores}) also requires the number of nodes in each class. Normalizing the affinity scores by the baseline scores results in the standard ratio scores, $h_{c, \alpha, t}^{\mathrm{VBK}1}$ (Eq.~\eqref{eq:vbk-standard-ratio-score}). Other scores can be computed analogously.
At hyperedge size $\alpha = 2$ (Fig.~\ref{fig:example-hypergraph}\textbf{(d)}), heterogeneous hyperedges ($t_c = 1$) are overrepresented and homogeneous hyperedges ($t_c = 2$) are underrepresented, irrespective of the reference class $c$. This means that the connection pattern at size $\alpha = 2$ exhibits heterophily. At hyperedge size $\alpha = 3$ (Fig.~\ref{fig:example-hypergraph}\textbf{(e)}) we find a more complicated mixing pattern: For reference class $c = 0$, homogeneous hyperedges are overrepresented, and heterogeneous hyperedges are underrepresented. In contrast, for reference class $c = 1$ homogeneous hyperedges are completely absent and heterogeneous edges are overrepresented. As the scores $h_{0, 3, t}^{\mathrm{VBK}1}$ are monotonically increasing for all $t\geq \lceil \frac{\alpha}{2} \rceil = 2$, class $0$ exhibits strict monotonic homophily. However, class $0$ does not exhibit strict majority homophily, but only simple homophily, because $h_{0, 3, 2}^{\mathrm{VBK}1}<1$ while $h_{0, 3,3}^{\mathrm{VBK}1}>1$. Class $1$ exhibits none of these notions of homophily. For hyperedge size $\alpha = 4$ (Fig.~\ref{fig:example-hypergraph}\textbf{(f)}), we observe that neither class exhibits majority nor monotonic homophily. On the contrary, homogeneous hyperedges are underrepresented, while hyperedges containing class $0$ and class $1$ nodes in equal proportions are most strongly overrepresented with respect to the baseline of either reference class. 

This example highlights an important feature of higher-order homophily: Mixing patterns can change with hyperedge size. Here, hyperedges of size $\alpha = 2$ and $\alpha = 4$ facilitate heterophilic interactions, whereas reference class $0$ exhibits homophily in interactions of size $\alpha = 3$.

\subsubsection{Further Considerations}

In the following, we highlight important properties of hypergraph homophily, including the combinatorial impossibility result of VBK. 

A desirable property of the standard affinity scores is that they are a proper generalization of the graph homophily index~\cite{coleman1958_homophily, currarini2009_economicmodelfriendship, altenburger2018_monophily}, i.e., both quantities coincide for $2$-uniform hypergraphs. Similarly, the normalized bias scores generalize the Coleman index~\cite{coleman1958_homophily} also known as inbreeding homophily index~\cite{currarini2009_economicmodelfriendship}.

For graphs, the homophily index can be statistically grounded in the stochastic block model (SBM)~\cite{altenburger2018_monophily}. Similarly, hypergraph homophily can be examined within the context of the cardinality-based hypergraph stochastic block model (CB-HSBM, Section~\ref{sec:hypergraph-sbm}), allowing the standard and alternative affinity scores to be understood as maximum-likelihood estimates of the success probability of a Binomial model or Poisson model, respectively~\cite{veldt2023_hypergraphhomophily}. Connecting homophily measures to specific generative models and statistical estimation is appealing, as it can facilitate identifying current shortcomings and opportunities for further developments (see e.g.,~\cite{altenburger2018_monophily}). 

One of the main challenges when using hypergraph homophily is that the measure is high-dimensional: At hyperedge size $\alpha$, each class $c$ is characterized by $\alpha$ numbers. In real-world datasets that feature hyperedges of different and potentially large sizes, this can make it hard to identify meaningful patterns.
The categories simple, (strict) majority, and (strict) monotonic homophily can offer some guidance but do not alleviate the fundamental problem.

To apply hypergraph homophily to hypergraphs with more than two classes, one needs to declare one class $c$ the reference class and treat all other classes as a uniform class of ``others''. This means hypergraph homophily measures the preference of nodes of the reference class $c$ to appear in hyperedges with a certain number of nodes of their own class, i.e., the type $t_c$, and is blind to the composition of hyperedges beyond that.

Finally, the combinatorial impossibility results of VBK establish that certain homophily patterns cannot be realized in hypergraphs~\cite{veldt2023_hypergraphhomophily}. At first glance, these results seem to pertain mainly to hypergraph generation, but they are also important for measurement, as they inform our expectations for the most extreme homophily patterns that are possible in any dataset. Assuming a hypergraph with nodes from two classes and that the standard baseline scores $b_{c, \alpha, t}^{\mathrm{VBK}1}$ are realizable, the following statements constrain strict monotonic and strict majority homophily: First, it is impossible for both classes to  exhibit strict monotonic homophily at odd hyperedge sizes. If additionally $h_{c, \alpha, \frac{\alpha}{2}}^{\mathrm{VBK}1}> h_{c, \alpha, \frac{\alpha}{2}-1}^{\mathrm{VBK}1}$ for one of the classes, this is also impossible at even hyperedge sizes. Second, it is impossible for both classes to exhibit strict majority homophily at odd hyperedge sizes. If the additional condition $h_{c, \alpha, \frac{\alpha}{2}}^{\mathrm{VBK}1}>1$ is satisfied, this is also true at even hyperedge sizes. VBK point out that similar statements can also be shown for alternative scores and normalized bias scores.

\subsection{Perplexity-Homophily Index}\label{sec:perplexity-homophily-index}

Kumar, Saxena, and Meena (KSM)~\cite{kumar2025_perplexityhomophilyindex} define a homophily measure, called the perplexity-homophily index, that quantifies the effective number of classes that participate on average in a hyperedge of a given hypergraph. In the remainder of this section, we introduce the perplexity-homophily index, show how to evaluate it with an example, and finally highlight important practical considerations.

\subsubsection{Definition}

The central object of the perplexity-homophily index is the \textit{interaction perplexity} $a^\mathrm{KSM}(e)$ of a hyperedge $e$, defined as 
\begin{equation}\label{eq:ksm-perplexity}
       \log_2 a^\mathrm{KSM}(e)
       =
       -\sum_{c\in\mathcal{C}} \frac{t_c(e)}{|e|}\log_2 \frac{t_c(e)}{|e|}
       .
\end{equation}
This means the interaction perplexity of a hyperedge is the exponentiated Shannon entropy of the class membership distribution of nodes within the hyperedge. Interaction perplexity is minimal ($a^\mathrm{KSM}(e)=1)$ in hyperedges that contain only nodes from one class, and maximal ($a^\mathrm{KSM}(e)=\min\{ \left|e \right|,C \}$) in hyperedges that contain the maximum number of classes in equal proportions. This is easy to see using the properties of the entropy of a categorical random variable with $C$ outcomes. Equation~\eqref{eq:ksm-perplexity} has a long history and is known as the Hill number of order $1$ or effective number of species in ecology~\cite{hill1973_diversity} and as perplexity or effective number of states in information theory~\cite{cover2006_elementsinformationtheory}. We note that an entropy-based approach to within-hyperedge diversity is also used in~\cite{guan2025_hypergraphcontrastivelearning}, however a proper null model is lacking.

To assess the magnitude of the interaction perplexity, the authors formulate a null model and compare $a^\mathrm{KSM}(e)$ to its expected value, called the \textit{baseline perplexity} $b_{\alpha}^{\mathrm{KSM}}$, under this null model. The baseline perplexity depends only on the hyperedge size $\alpha = |e|$. Under the null model, random hyperedges of size $\alpha$ are created by drawing $\alpha$ nodes without replacement from the node set $\mathcal{V}$ with probability proportional to their $\alpha$-degree $d_\alpha(i)$ (Eq.~\eqref{eq:degree-size-dependent}) but independent of their class membership $c_i$. The authors do not provide a closed-form expression for the baseline perplexity, $b_\alpha^{\mathrm{KSM}}$, but instead approximate its values using Monte Carlo sampling. 

The comparison of a hyperedge's diversity to the baseline perplexity yields the \textit{normalized diversity gap},
\begin{equation}
    h^\mathrm{KSM}(e)
    =
    \frac{
            b_{|e|}^{\mathrm{KSM}}-a^{\mathrm{KSM}}(e)
        }{
            b_{|e|}^{\mathrm{KSM}}-1
        }
        ,
\end{equation}
where $b_{|e|}^{\mathrm{KSM}}-1$ is the maximum value $b_{|e|}^{\mathrm{KSM}}-a^{\mathrm{KSM}}(e)$ can take.

The normalized diversity gap is a hyperedge-level quantity and can be aggregated to obtain a hypergraph-level quantity or a hyperedge size-dependent quantity. The authors choose aggregation by averaging to obtain the \textit{perplexity homophily index},
\begin{equation}
    h^{\mathrm{KSM}}
    =
    \frac{1}{E} \sum_{e\in \mathcal{E}} h^\mathrm{KSM}(e),
\end{equation}
or, alternatively, a hyperedge size-dependent quantity, that we refer to as the \textit{hyperedge size-dependent perplexity homophily index},
\begin{equation}
    h^\mathrm{KSM}_{\alpha}
    =
    \frac{1}{E_\alpha} \sum_{e\in \mathcal{E}} h^\mathrm{KSM}(e)
    \,
    \mathbf{1} \{ |e| = \alpha\}.
\end{equation}

\subsubsection{Example}

\begin{figure}
    \centering
    \includegraphics[width=0.99\linewidth]{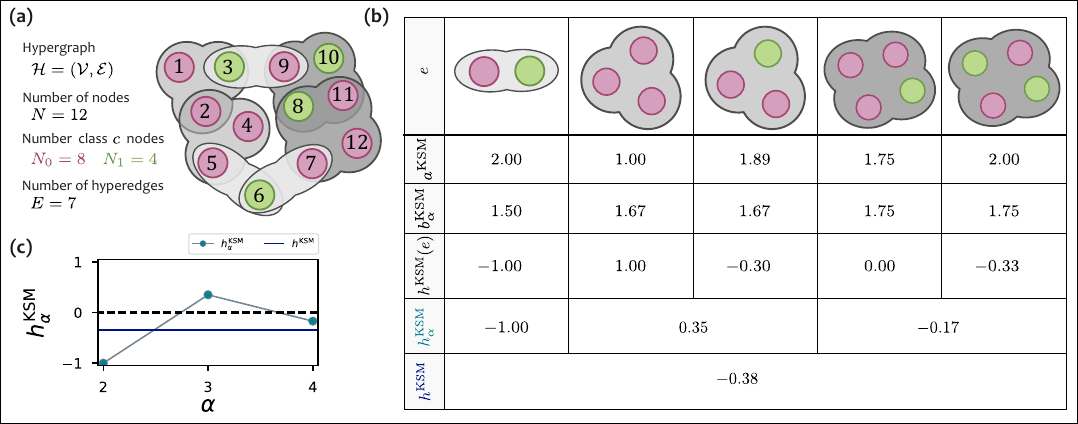}
    \caption{\textbf{Perplexity-Homophily Index:} \textbf{(a)} In the example hypergraph $\mathcal{H}$, $N=12$ nodes are divided into two classes. $N_0=8$ nodes are members of class $0$ (purple) whereas $N_1=4$ of them are members of class $1$ (green). The total number of hyperedges is $E=7$. \textbf{(b)} The perplexity-homophily index $h^\mathrm{KSM}$ (dark blue) and its hyperedge size-dependent version $h^\mathrm{KSM}_\alpha$ (light blue) are computed from the interaction perplexity $a^{\mathrm{KSM}}(e)$ of each edge compared to the hyperedge size-dependent baseline $b^{\mathrm{KSM}}_\alpha$ (using $10^5$ Monte Carlo samples). \textbf{(c)} Inspecting $h^\mathrm{KSM}_\alpha$ (light blue circles) as a function of the hyperedge size $\alpha$ compared to the neutral baseline $h^\mathrm{KSM}_\alpha = 0$ (dashed black line) shows that the homophily varies with hyperedge size $\alpha$, with homophily at $\alpha = 3$ but heterophily for $\alpha \in \{2, 4\}$. The perplexity-homophily index of the entire hypergraph indicates heterophily $h^\mathrm{KSM} = -0.38$ (dark blue line).}
    \label{fig:example-perplexity}
\end{figure}

Here, we show how to calculate the perplexity-homophily index for the hypergraph $\mathcal{H}$ in Fig.~\ref{fig:example-perplexity}\textbf{(a)}. The hypergraph has $N=12$ nodes that are members of one of two classes $c\in \{0,1\}$ consisting of $N_0=8$ and $N_1=4$ nodes, respectively. In Fig.~\ref{fig:example-perplexity}\textbf{(b)}, we calculate the interaction perplexity $a^{\mathrm{KSM}}(e)$ for several example hyperedges $e$, and compare its value to the baseline perplexity $b_{|e|}^ {\mathrm{KSM}}$ determined using $10^5$ Monte Carlo samples. This yields the normalized diversity gap, $h^{\mathrm{KSM}}(e)$. The perplexity-homophily index $h^\mathrm{KSM}$ is obtained by averaging the normalized diversity gap over all hyperedges. The hyperedge size-dependent version $h^\mathrm{KSM}_\alpha$ restricts averages to hyperedges of a fixed size $\alpha$. Inspecting the latter, in Fig.~\ref{fig:example-perplexity}\textbf{(c)}, as a function of the hyperedge size $\alpha$ in comparison to the neutral baseline $h_{\alpha}^\mathrm{KSM} = 0$ reveals homophily for edges of size $\alpha = 3$ but heterophily in hyperedges of size $\alpha \in \{2, 4\}$. The perplexity-homophily index of the entire hypergraph is $h^\mathrm{KSM} = -0.38$. This means the measure classifies the hypergraph as overall heterophilous.

\subsubsection{Further Considerations}

The perplexity-homophily index puts hyperedges front and center by assessing how diverse members of a single hyperedge are, and moving from this microscopic measurement to meso- and macroscale summary statistics to reduce complexity. The measure also naturally deals with more than two classes. Through its relationship to ecology, this approach also invites researchers to explore interdisciplinary perspectives on quantifying higher-order mixing patterns.

An unexpected property of the perplexity-homophily index is that it closely approximates Newman's nominal assortativity coefficient~\cite{newman2003_mixingpatterns, kumar2025_perplexityhomophilyindex} for $\alpha = 2$---a fact that KSM state without proof, but that can be verified by observing that the probability of sampling a same-class pair under KSM's degree-biased node sampling baseline is approximately equal to the probability of a same-class pair under Newman's stub-mixing null model.

The size-dependent perplexity homophily index is especially attractive from a practical perspective, as it condenses information about different types of hyperedges of the same size into a single number while still allowing one to track changes in homophily with varying hyperedge size. 

The perplexity-homophily index's null model explicitly accounts for degree heterogeneity, but sacrifices analytical tractability. The baseline perplexity is determined via Monte Carlo sampling, making it necessary to determine the appropriate sample size for a given hypergraph. Only in the special case that all nodes within a class have the same degree, can the null model be related to Wallenius' noncentral hypergeometric distribution, but this assumption is unlikely to be met in practice~\cite{fog2008_calculationmethodswallenius}.

\subsection{Simplicial Homophily}\label{sec:simplicial-homophily}

Simplicial homophily~\cite{sarker2023_simplicialconference, sarker2024_simplicialpnas}, introduced by Sarker, Northrup, and Jadbabaie (SNJ), is specifically designed for fully-nested hypergraphs (also called simplicial complexes). This means, e.g., if three nodes interact in a hyperedge of size $3$, all pairwise interactions between them also occur. This is sometimes called downward-closure. Fully-nested hypergraphs are often used as models of face-to-face interactions~\cite{battiston2025_collectivehumanbehavior} and in mathematical models of complex contagion~\cite{majhi2022_dynamicshigherorder, battiston2026_collectivedynamics}.

If it is known, a priori, that hypergraphs in a given application are fully nested, simplicial homophily offers a way to account for this prior knowledge through an appropriate null model. It answers the question: How homophilic are interactions encoded in hyperedges of a certain size beyond what one would expect given the connections at smaller hyperedge sizes?

In this section, we first define the simplicial homophily measures of SNJ, illustrate through an example how they answer the above question, and comment on important considerations when using simplicial homophily in practice.

\subsubsection{Definition}

The definition of simplicial homophily mirrors that of hypergraph homophily (Section~\ref{sec:hypergraph-homophily}): First, affinity scores are defined, which are then normalized by baseline scores from an appropriate null model.
However, simplicial homophily is only applicable to hypergraphs with fully-nested hyperedges. In this section, we focus on \textit{class-independent simplicial homophily} $h^{\mathrm{SNJ}1}_\alpha$ and only briefly introduce the class-dependent version $h^{\mathrm{SNJ}2}_{c, \alpha, t}$ at the end of this section.
The \textit{class-independent affinity scores} $a^{\mathrm{SNJ}1}_{\alpha}$ for hyperedges of size $\alpha$ are defined as the fraction of homogeneous hyperedges of this size,
\begin{equation}\label{eq:snj-affinity-score}
    a_{\alpha}^{\mathrm{SNJ1}}
    =
    \frac{
        \sum_{c \in \mathcal{C}} E_{\alpha,\alpha}^{(c)}
    }{
        E_\alpha
    }
    .
\end{equation}
The \textit{class-independent simplicial baseline scores} $b_{\alpha}^\mathrm{SNJ1}$ at hyperedge size $\alpha$ are based on a null model that explicitly incorporates the assumption of a fully nested hypergraph: Starting from a hypergraph $\mathcal{H}$, imagine the hypergraph $\mathcal{H}^{(<\alpha)}$ consisting of hyperedges up to size $\alpha - 1$, and then add all hyperedges of size $\alpha$ that are compatible with the $\alpha - 1$ interactions, obtaining a hypergraph $\overline{\mathcal{H}^{(<\alpha)}}$, called the closure of $\mathcal{H}^{(<\alpha)}$. The class-independent simplicial baseline scores at size $\alpha$ are the class-independent simplicial affinity score of the hypergraph $\overline{\mathcal{H}^{(<\alpha)}}$,
\begin{equation}\label{eq:snj-simplicial-baseline}
    b_{\alpha}^{\mathrm{SNJ1}}(\mathcal{H})
    =
    a_{\alpha}^{\mathrm{SNJ}1}(\overline{\mathcal{H}^{(<\alpha)}})
    ,
\end{equation}
where we make the dependence on the underlying hypergraph explicit.
The baseline score $b_{\alpha}^\mathrm{SNJ1}$ has a clear probabilistic interpretation: It is the probability that a hyperedge $e$ of size $\alpha$ is homogeneous if placed uniformly at random among the permissible locations in $\mathcal{H}^{(<\alpha)}$.

The \textit{class-independent simplicial homophily} $h_{\alpha}^{\mathrm{SNJ}1}$ is defined as a ratio score,
\begin{equation}\label{eq:snj-simplicial-homophily}
    h_{\alpha}^{\mathrm{SNJ}1}
    =
    \frac{
        a_{\alpha}^{\mathrm{SNJ}1}
        }{
        b_{\alpha}^{\mathrm{SNJ}1}
        }
    .
\end{equation}
A main goal of SNJ is to question the validity of VBK's null model if hypergraphs are fully nested. To this end, they introduce a \textit{class-independent hypergraph homophily baseline score}, $b^{\mathrm{SNJ}\mathcal{H}}_\alpha$ as
\begin{equation}\label{eq:snj-hypergraph-baseline}
    b_{\alpha}^{\mathrm{SNJ}\mathcal{H}}
    =
    \frac{
        \sum_{c \in \mathcal{C}}\binom{N_c}{\alpha}
    }{
        \binom{N}{\alpha}
    }
    ;
\end{equation}
this is the probability to obtain a homogeneous hyperedge when drawing $\alpha$ nodes uniformly at random from the set of class-labeled nodes, thereby creating a hyperedge of size $\alpha$ that may not conform with the constraint of a fully-nested hypergraph. The \textit{class-independent hypergraph homophily} is then defined as,
\begin{equation}\label{eq:snj-hypergraph-homophily}
    h_{\alpha}^{\mathrm{SNJ}\mathcal{H}}
    =
    \frac{a^{\mathrm{SNJ}1}_\alpha}{b_{\alpha}^{\mathrm{SNJ}\mathcal{H}}}
    .
\end{equation}
We emphasize that the class-independent hypergraph homophily, Eq.~\eqref{eq:snj-hypergraph-homophily}, is distinct from VBK's hypergraph homophily ratio score---in both the standard, Eq.~\eqref{eq:vbk-standard-ratio-score}, and alternative version, Eq.~\eqref{eq:vbk-alternative-ratio-score}---as VBK's scores are class-dependent and only consider hyperedges that contain at least one node of the reference class when computing affinity and baseline scores.
SNJ show that Eq.~\eqref{eq:snj-simplicial-baseline} is a more appropriate null model for fully-nested hypergraphs than Eq.~\eqref{eq:snj-hypergraph-baseline} in the following sense: Even if hyperedges of size $\alpha$ are inserted uniformly at random, i.e., without any preference for class composition, into a fully-nested hypergraph of maximum hyperedge size $\alpha - 1$ such that the hypergraph remains fully nested, Eq.~\eqref{eq:snj-hypergraph-homophily} will indicate homophilous interactions at hyperedge size $\alpha$ if interactions at hyperedge size $\alpha - 1$ are homophilous. Equation~\eqref{eq:snj-simplicial-homophily} on the other hand will classify these interactions as neutral. This statement can be made precise in the context of the simplicial stochastic block model~\cite{sarker2024_simplicialpnas}.

For completeness, we state the definitions of \textit{class-dependent simplicial affinity scores} $a_{c, \alpha, t}^{\mathrm{SNJ}2}$, \textit{baseline scores} $b_{c, \alpha, t}^{\mathrm{SNJ}2}$, and \textit{homophily} $h_{c, \alpha, t}^{\mathrm{SNJ}2}$, which are explicitly based on the standard affinity and baseline scores of VBK and allow us to quantify homophily at hyperedge size $\alpha$ and type $t$ with respect to a reference class $c$,
\begin{align}
\begin{split}
    a_{c, \alpha, t}^{\mathrm{SNJ}2}(\mathcal{H})
    &=
    a_{c, \alpha, t}^{\mathrm{VBK}1}(\mathcal{H}^{(\alpha)}) \\ 
    b_{c,\alpha,t}^{\mathrm{SNJ}2}(\mathcal{H})
    &=
    a_{c, \alpha, t}^{\mathrm{VBK}1}(\overline{\mathcal{H}^{(<\alpha)}}) \\
    h_{c, \alpha, t}^{\mathrm{SNJ}2}(\mathcal{H})
    &=
    \frac{a_{c, \alpha, t}^{\mathrm{SNJ}2}(\mathcal{H})}{b_{c, \alpha, t}^{\mathrm{SNJ}2}(\mathcal{H})}
    .
\end{split}
\end{align}
SNJ conjecture that this class-wise definition induces similar combinatorial impossibility results as VBK's hypergraph homophily, and thus focus on class-independent measures.

\subsubsection{Example}

\begin{figure}
    \centering
    \includegraphics[width=0.99\linewidth]{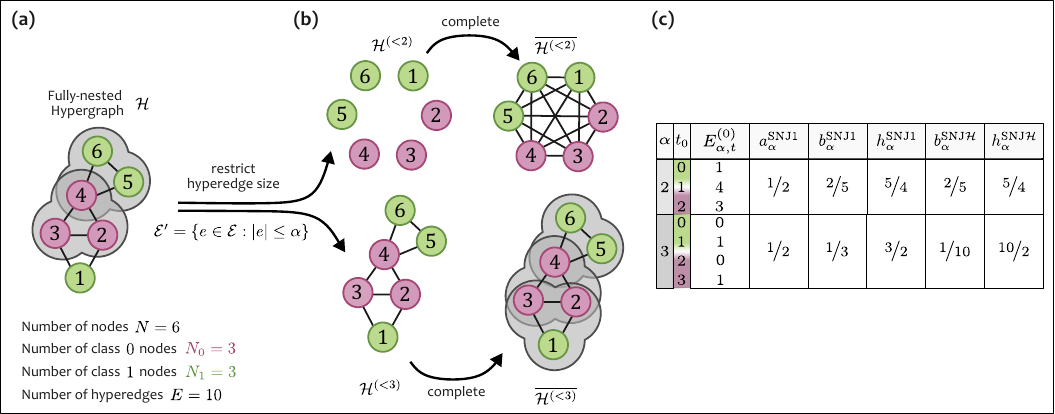}
    \caption{\textbf{Simplicial Homophily:} \textbf{(a)} A fully-nested hypergraph $\mathcal{H}$ with $E=10$ hyperedges of size $\alpha \in \{2,3\}$ and $N=6$ nodes, $N_0=3$ of which belong to class $0$ (purple) and $N_1=3$ of which belong to class $1$ (green). \textbf{(b)} To compute the class-independent simplicial baseline scores, auxiliary hypergraphs $\mathcal{H}^{(<\alpha)}$ containing only hyperedges of size $\alpha'< \alpha$ need to be constructed. Then, their completion $\overline{\mathcal{H}^{(<\alpha)}}$ needs to be determined. \textbf{(c)} The class-independent simplicial affinity scores $a_\alpha^{\mathrm{SNJ}1}$, the simplicial baseline scores $b_{\alpha}^{\mathrm{SNJ}1}$, and hypergraph baseline $b_{\alpha}^{\mathrm{SNJ}\mathcal{H}}$ scores, as well as the respective homophily scores $h_\alpha^{\mathrm{SNJ}1}$ and $h_\alpha^{\mathrm{SNJ}\mathcal{H}}$, are computed from the hyperedge counts $E_{\alpha, t}^{(0)}$. While the simplicial homophily $h^{\mathrm{SNJ1}}_{\alpha}$ and hypergraph homophily $h^{\mathrm{SNJ\mathcal{H}}}_{\alpha}$ scores are, by definition, equal for $\alpha = 2$, the simplicial homophily is much lower than the hypergraph homophily at $\alpha = 3$ as the simplicial null model accounts for the constraint that the hypergraph needs to be fully nested.}
    \label{fig:example-simplicial}
\end{figure}

To illustrate the definition of class-independent simplicial homophily, we start from the fully-nested hypergraph $\mathcal{H}$ in Fig.~\ref{fig:example-simplicial}\textbf{(a)}. The hypergraph has $N=6$ nodes that come in equal proportions from class $0$ (purple, $N_0=3$) and class $1$ (green, $N_1=3$). The total number of hyperedges is $E=10$. The auxiliary hypergraphs $\mathcal{H}^{(<\alpha)}$ for hyperedges of size $\alpha = 2$ and $\alpha = 3$ are constructed by restricting the set of hyperedges to those of size less than $\alpha$ (Fig.~\ref{fig:example-simplicial}\textbf{(b)}). In this construction each node is regarded as part of its own $\alpha = 1$ hyperedge. Subsequently, the completion $\overline{\mathcal{H}^{(<\alpha)}}$ of $\mathcal{H}^{(<\alpha)}$ is formed, i.e., all hyperedges of size $\alpha$ that are compatible with the constraint of a fully-nested hypergraph are inserted. For $\alpha = 2$, this always results in a complete graph. To compute the class-independent simplicial homophily, we determine the total number $E_\alpha$ of hyperedges of size $\alpha$ and the number of homogeneous hyperedges with respect to each class. In this example with two classes, this is equivalent to $E^{(0)}_{\alpha, 0}$ and $E^{(0)}_{\alpha, \alpha}$ with respect to the arbitrarily chosen reference class $c=0$. The class-independent simplicial affinity (Eq.~\eqref{eq:snj-affinity-score}) and baseline (Eq.~\eqref{eq:snj-simplicial-baseline}) scores are the fraction of homogeneous hyperedges in $\mathcal{H}$ and $\overline{\mathcal{H}^{(<\alpha)}}$, respectively. The class-independent hypergraph baseline scores are computed using Eq.~\eqref{eq:snj-hypergraph-baseline}. The class-independent simplicial and class-independent hypergraph homophily scores are determined as the ratio of affinity scores to baseline scores. For hyperedge size $\alpha = 2$, the class-independent simplicial homophily, $h^{\mathrm{SNJ}1}_2$, and hypergraph homophily, $h^{\mathrm{SNJ}\mathcal{H}}_2$, are, by construction, equal $h^{\mathrm{SNJ}1}_{2}=h^{\mathrm{SNJ}\mathcal{H}}_2 = 5/4$ and indicate that homophilous interactions are more abundant than expected under the baseline. For hyperedges of size $\alpha = 3$, the class-independent simplicial homophily, $h^{\mathrm{SNJ}1}_3 =3/2$, is much lower than the class-independent hypergraph homophily, $h^{\mathrm{SNJ}\mathcal{H}}_3 = 10/2$, as the null model of the former takes into account that $\alpha = 3$ hyperedges need to obey the constraints imposed by the already homophilous sub-hypergraph $\mathcal{H}^{(<3)}$.

\subsubsection{Further Considerations}

We conclude this section on simplicial homophily by commenting on important considerations when using the measure for data analysis or further theoretical work.

Simplicial homophily is only defined for fully-nested hypergraphs with the goal of accounting for the combinatorial constraints imposed by the downward closure property. We strongly advocate against applying simplicial homophily to general hypergraphs, in which downward closure is artificially enforced by inserting or deleting hyperedges, as such manipulations unnecessarily distort the data.

When deciding whether to use simplicial homophily and when interpreting the measure, it is important to understand how precisely hyperedges of different sizes are treated differently: Larger hyperedges are generated conditioned on smaller hyperedges not the other way around. If this ``bottom-up'' process is a reasonable null assumption, simplicial homophily can be an adequate measure. Such a process could, e.g., describe a social gathering in which people meet in small groups and these small groups coalesce into larger groups over time. If, on the other hand, a ``top-down'' process that inserts smaller hyperedges among the members of larger hyperedges is a better description, then simplicial homophily might not be the appropriate choice. An example for such a situation could be a classroom setting in which large groups are assigned by a teacher and pairwise interactions take place within these groups.

The class-independent version of the simplicial homophily yields a single number $h_\alpha^{\mathrm{SNJ}1}$ per hyperedge size independent of hyperedge type or the number of classes. On the one hand, this makes the measure easier to visualize and inspect. On the other hand, it neglects potentially meaningful differences between classes as well as most of the information in the type dependence of the hyperedge counts, $E^{(c)}_{\alpha, t}$. Taking only homogeneous hyperedges, i.e., the counts $E_{\alpha, \alpha}^{(c)}$, into account can be problematic for large hyperedge sizes $\alpha$, for which perfectly homogeneous hyperedges are potentially rare. Class-dependent simplicial homophily addresses these problems but remains less explored.

From a computational perspective, it is worth pointing out that the simplicial baseline scores depend on the exact hypergraph structure, not just summary statistics of the data. Evaluating the baseline scores requires access to libraries that permit manipulating attributed hypergraphs to compute the completion, $\overline{\mathcal{H}^{(<\alpha)}}$, and to count the number of homogeneous hyperedges of a given size. 

\subsection{Message Passing Homophily}\label{sec:message-passing-homophily}

Outside of the social sciences, homophily has become a central theme in the graph machine learning literature, as a key factor shaping the performance of predictive models. As machine learning architectures are adapted to hypergraphs, it is natural to seek analogues of homophily in the higher-order setting.

In this section, we explore one such extension, the message passing homophily of Telyatnikov et al.~\cite{telyatnikov2024_mpnnhomophily}, provide an example computation, and discuss its prospects and limitations.

\subsubsection{Definition}

\textit{Message passing homophily}~\cite{telyatnikov2024_mpnnhomophily} takes inspiration from the design of message passing neural networks (MPNNs) for hypergraphs. These machine learning architectures start from a vector of initial node or hyperedge attributes and refine them through an iterative message passing process to arrive at embedding vectors for nodes or hyperedges that are used for predictive tasks.

Similarly, the definition of message passing homophily starts from an initial homophily value, which is iteratively refined over $\tau_\mathrm{max}$ iterations of message passing (MP). The initial \textit{hyperedge-level MP homophily score} $a^{\mathcal{E}\mathrm{MP}}_{c, 0}(e)$ with respect to class $c$ is the fraction of nodes of class $c$ in the hyperedge $e$,
\begin{equation}\label{eq:mpnn-0-edge}
    a^{\mathcal{E}\mathrm{MP}}_{c, 0}(e)
    =
    \frac{t_c(e)}{|e|},
\end{equation}
where $t_c(e)$ is $e$'s type with respect to class $c$ (Eq.~\eqref{eq:type}). The subsequent iterates of the hyperedge-level MP homophily score and node-level MP homophily score are defined via a two-step message passing process. First, the \textit{node-level MP homophily score} $a^{\mathcal{V}\mathrm{MP}}_{\tau}(i)$ of each node $i$ at iteration $\tau$ is defined based on the hyperedge-level scores at the same iteration $\tau$,
\begin{equation}\label{eq:mpnn-tau-node}
    a^{\mathcal{V}\mathrm{MP}}_{\tau}(i)
    =
    \phi_\mathcal{V}(
        \{\!\{ 
            a^{\mathcal{E}\mathrm{MP}}_{c_i,\tau}(e) 
        \}\!\}_{
                e\in \mathcal{E}\,:\,i\in e
                }
    )
    ,
\end{equation}
where $\{\!\{ \cdot\}\!\}$ denotes multisets, and $\phi_\mathcal{V}$ is called an aggregation function and maps multisets of real numbers to real numbers, e.g., via averaging. Note that only the current hyperedge-level MP homophily scores of hyperedges of which node $i$ is a member influence its node-level MP homophily score. We call this step the $\mathcal{V}$-step.

Second, the iteration-$\tau$ \textit{hyperedge-level MP homophily score} $a^{\mathcal{E}\mathrm{MP}}_{c,\tau}(e)$ with respect to class $c$ for a hyperedge $e$ is defined based on the node-level MP homophily scores of its members,
\begin{equation}\label{eq:mpnn-tau-edge}
    a^{\mathcal{E}\mathrm{MP}}_{c, \tau+1}(e)
    =
    \phi_\mathcal{E}(
        \{\!\{
        a^{\mathcal{V}\mathrm{MP}}_{\tau}(i)
        \}\!\}_{
                i \in e \,:\, c_i=c
                }
    )
    ,
\end{equation}
where $\phi_\mathcal{E}$ is an aggregation function, potentially chosen differently from $\phi_\mathcal{V}$. We call this step the $\mathcal{E}$-step. We note that similar quantities have previously been used, however without iterative refinement~\cite{chun2025_NoAH, li2025_heterophily}.

In addition to the node- and hyperedge-level MP homophily scores, Telyatnikov et al. introduce the $\Delta$-\textit{homophily} at iteration $\tau$ as the fraction of nodes for which the change in node-level MP homophily with respect to the previous iteration is bounded by $\mu > 0$,
\begin{equation}\label{eq:mpnn-delta-homophily}
    a^{\Delta \mathrm{MP}}_\tau(\mu)
    =
    \frac{1}{N} \sum_{i \in \mathcal{V}} \mathbf{1}
        \left\{
        \left| 
            a^{\mathcal{V}\mathrm{MP}}_{\tau}(i) - a^{\mathcal{V}\mathrm{MP}}_{\tau-1}(i)
        \right|
        <
        \mu
        \right\}.
\end{equation}

It is important to realize that none of these quantities are compared to their value under a null model.

\subsubsection{Example}

\begin{figure}
    \centering
    \includegraphics[width=0.99\linewidth]{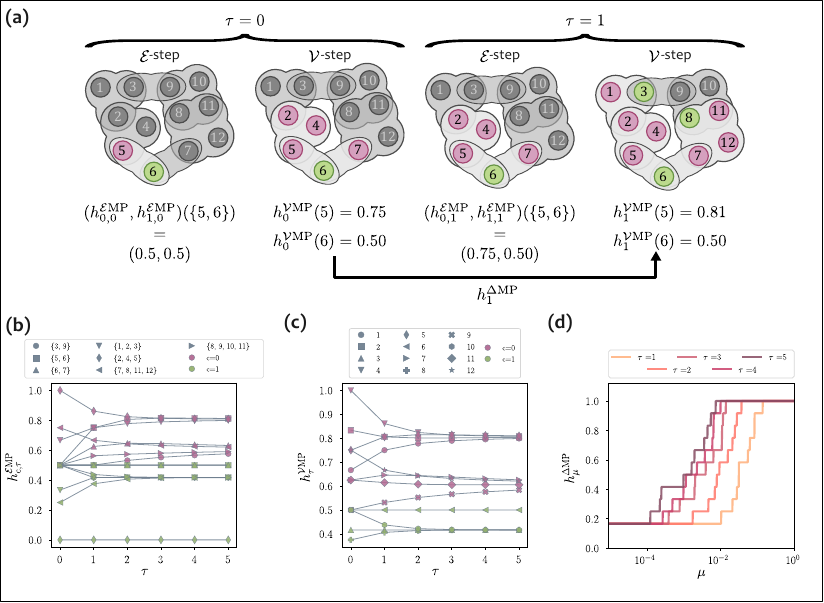}
    \caption{\textbf{Message Passing Homophily:} \textbf{(a)} Hyperedge-level MP homophily $a^{\mathcal{E}\mathrm{MP}}_{c,\tau}(\{5,6\})$ and node-level MP homophily $a^{\mathcal{V}\mathrm{MP}}_{\tau}(i)$ for nodes $i \in \{5, 6\}$ are calculated via $\tau$ steps of message passing. Nodes from class $0$ (purple) and class $1$ (green) are shown in color at step $\tau$ if they influence the value of these quantities and in gray otherwise. The $\Delta$-homophily, $a^{\Delta \mathrm{MP}}_\tau(\mu)$, is the fraction of nodes for which the change between $a^{\mathcal{V}\mathrm{MP}}_{\tau}(i)$ and $a^\mathrm{\mathcal{V}\mathrm{MP}}_{\tau -1}(i)$ is bounded by $\mu$. \textbf{(b)} The hyperedge-level message passing homophily, $h^{\mathcal{E}\mathrm{MP}}_{c,\tau}$ with respect to class $c=0$ (purple) and the class $c=1$ (green) as a function of the iteration $\tau$ for the $E=7$ different hyperedges stabilizes. \textbf{(c)} The node-level message passing homophily $h^{\mathrm{\mathcal{V}\mathrm{MP}}}_{\tau}(i)$ stabilizes as the iterations $\tau$ for nodes in class $c=0$ (purple) as well as class $c=1$ (green). \textbf{(d)} The $\Delta$-homophily $h^{\Delta \mathrm{MP}}_{\tau}(\mu)$ increases at lower values of $\mu$ and plateaus earlier as the number of iterations $\tau$ increases. This means that there is on average less change in $h^{\mathcal{V}\mathrm{MP}}_{\tau}(i)$ at later iterations.
    }
    \label{fig:example-message-passing}
\end{figure}

Figure~\ref{fig:example-message-passing}\textbf{(a)} shows how to compute MP homophily for a small hypergraph with $N=12$ nodes and $E=7$ hyperedges, which belong to either class $0$ (purple) or class $1$ (green). We focus on the hyperedge $e = \{5,6\}$ for computing hyperedge-level MP homophily and on its members for computing node-level MP homophily. Only parts of the hypergraph that influence the value of the MP homophily of $e$ or its members at a given iteration are shown in color. In the $\mathcal{E}$-step of iteration $\tau = 0$, the hyperedge-level MP homophily of $e$ with respect to either class $c$ is $a_{c,0}^{\mathcal{E}\mathrm{MP}}=0.5$, as $e$ contains one member of each class. Only $e$ itself influences $e$'s homophily in this step. In the $\mathcal{V}$-step of iteration $\tau=0$, the node-level MP homophily of $e$'s members $5$ and $6$ is updated based on the hyperedge-level MP homophily with respect to their respective classes $c_5=0$ and $c_6=1$. In the next iteration's ($\tau=1$) $\mathcal{E}$-step, $e$'s hyperedge-level MP homophily changes in response to changes in the node-level MP homophily of its members, thereby incorporating information about the class composition of adjacent hyperedges. In the $\mathcal{V}$-step, node $5$'s node-level MP homophily changes influenced by the changes in node-level MP homophily of adjacent nodes of the same class. Through these changes node $5$'s homophily incorporates information from its $1$-hop neighborhood. In contrast, node $6$'s homophily does not change as it has no immediate neighbors belonging to the same class. Finally, the $\Delta$-homophily summarizes global changes in node-level MP homophily. Figure~\ref{fig:example-message-passing}\textbf{(b)} displays how the hyperedge-level MP homophily $h^{\mathcal{E}\mathrm{MP}}_{c,\tau}(e)$ of each edge $e$ with respect to class $c=0$ (purple) and class $c=1$ (green) stabilizes as the number of iterations $\tau$ increases. The node-level MP homophily $h_{\tau}^ {\mathcal{V}\mathrm{MP}}(i)$ (Fig.~\ref{fig:example-message-passing}\textbf{(c)}) of individual nodes $i$ also stabilizes as the number of iterations $\tau$ increases. Nodes of class $0$ (purple) reach higher scores than nodes of class $1$ (green). The $\Delta$-homophily (Fig.~\ref{fig:example-message-passing}\textbf{(d)}) is bound to monotonically increase as a function of $\mu$, the increase starts and plateaus earlier in later iterations, meaning that the node-level MP homophily scores stabilize. It is important to be aware of the challenges that the lack of baseline or null model causes for interpreting MP homophily, e.g., it is not clear to what extent the higher node-level MP homophily scores of class-$0$ nodes are due to the hypergraphs structure or the higher abundance of class-$0$ nodes. 

\subsubsection{Further Considerations}

MP homophily differs from other homophily measures in that it defines a sequence of homophily scores that capture local neighborhoods of increasing size. While MP homophily is clearly tailored to applications in machine learning, the idea of incorporating a dynamical process into homophily measures could be of interest more broadly (see, e.g., Section~\ref{sec:RWHS}).

Furthermore, the iterative nature of $\Delta$-homophily makes it possible to inspect not only the absolute value of the homophily scores but also their change over iterations.
From a practical perspective, three aspects are worth noting: First, MP homophily is not a single measure but unlocks an entire design space by leaving the choice of aggregation function $\phi_\mathcal{V}$ in Eq.~\eqref{eq:mpnn-tau-node} and $\phi_\mathcal{E}$ in Eq.~\eqref{eq:mpnn-tau-edge} open. Any particular choice leads to a new measure with slightly different interpretation. Second, in contrast to other measures MP homophily cannot be computed from summary statistics like $(c, \alpha, t)$-hyperedge counts, but requires access to the structure of the hypergraph, preferably in a data structure that supports efficient implementation of message passing updates. Finally, Telyatnikov et al. do not specify any null model. This makes it difficult to interpret the obtained values even for nodes within a single hypergraph, let alone across hypergraphs. While comparing MP homophily scores to their values in several randomized versions of the same hypergraph (e.g., sampled from one of the null models in Section~\ref{sec:null models}) is conceptually simple, it is also computationally more expensive than evaluating closed-form expressions for baseline scores.

\subsection{Random Walk HyperSegregation}\label{sec:RWHS}

The \textit{Random Walk HyperSegregation} (RWHS)~\cite{failla2025_RHWS} defines, similar to the MP homophily (Section~\ref{sec:message-passing-homophily}), a family of homophily measures that leverage dynamical processes---in this case a random walk.

\subsubsection{Definition}

Failla, Rossetti, and Cauteruccio (FRC) take random walks on a hypergraph as a starting point for defining RWHS. The measure is agnostic to the exact definition of the random walk, as long as a realization of the random walk can be used to obtain sequences of nodes $\mathcal{S}_r=(i_1,\dots,i_T)$, where $T$ is called the length of the random walk and $r\in \{1,\dots, R\}$ enumerates different realizations of the random walk starting from the same node $i_0 \in \mathcal{V}$. Note that we exclude $i_0$ from $\mathcal{S}_r$.

For each node, FRC define the \textit{meet-wise RWHS} $a_{T,R}^{\mathrm{mRWHS}}(i_0)$ of node $i_0$ as the fraction of nodes in the same class as node $i_0$ along the length $T$ random walks and across the $R$ realizations 
\begin{equation}\label{eq:meet-RWHS}
    a_{T,R}^{\mathrm{mRWHS}}(i_0)
    =
    \frac{1}{R}
    \sum_{r=1}^R
    \frac{
        \left|
            \{
                i_t \in \mathcal S_r \;
                : \;
                c_{i_t} = c_{i_0}
            \}
        \right|
        }{
        T    
    }.
\end{equation}
Similarly, the \textit{jump-wise RWHS} $a^{\mathrm{jRWHS}}_{T,R}(i_0)$ of node $i_0$ is the fraction of subsequent nodes that are in the same class (as each other, not necessarily as $i_0$), i.e.,
\begin{equation}\label{eq:jump-RWHS}
        a_{T,R}^{\mathrm{jRWHS}}(i_0)
    =
    \frac{1}{R}
    \sum_{r=1}^R
    \frac{
        \left|
            \{
                i_{t} \in \mathcal{S}'_r \;
                : \;
                c_{i_{t+1}} = c_{i_t}
            \}
        \right|
        }{
        T-1    
    },
\end{equation}
where $\mathcal{S}'_r=\{i_t\in\mathcal{S}_r\;:\; t<T \}$ removes the last element from $\mathcal{S}_r$. FRC treat node-wise and jump-wise RWHS as node-wise measures and inspect their distribution in a given hypergraph, instead of constructing aggregate scores.

The meet-wise and jump-wise RWHS play the role of affinity scores and thus need to be compared to their values in a suitable null model. FRC do not provide closed form expressions for the value of meet-wise or jump-wise RWHS in any null model. However, they suggest comparing these measures to their values in a hypergraph randomized using a version of the Chung-Lu model for bipartite networks~\cite{aksoy2017_bipartitecl}. However, many other null models for hypergraphs could be used (see, e.g., Section~\ref{sec:null models}).

\subsubsection{Example}

\begin{figure}
    \centering
    \includegraphics[width=0.99\linewidth]{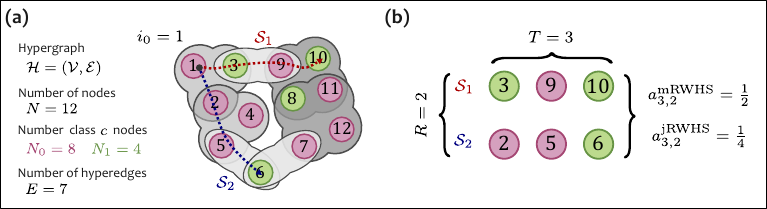}
    \caption{\textbf{Random Walk HyperSegregation:} \textbf{(a)} Starting from node $i_0=1$, two random walk realizations $\mathcal{S}_1$ (red) and $\mathcal{S}_2$ (blue) of length $T=3$ explore the example hypergraph $\mathcal{H}$ with $N=12$ nodes with $N_0=8$ from class $0$ (purple) and $N_1=4$ from class $1$. The total number of hyperedges is $E=7$. \textbf{(b)} Based on these $R=2$ realizations of length $T=3$ the meet-wise RWHS is computed as the fraction of nodes in the same class as the starting node in either realization, $a_{3,2}^{\mathrm{mRWHS}}=\frac{1}{2}$. The jump-wise RWHS is the fraction of subsequent nodes in either realization with the same class, $a_{3,2}^{\mathrm{jRWHS}}=\frac{1}{4}$.}
    \label{fig:example-RWHS}
\end{figure}

Figure~\ref{fig:example-RWHS}\textbf{(a)} shows two ($R=2$) random walk realizations $\mathcal{S}_1=(3, 9, 10)$ (red) and $\mathcal{S}_2=(2, 5,6)$ (blue) of length $T=3$ on a hypergraph $\mathcal{H}$ with $N=12$ nodes that are split among class $0$ (purple, $N_0=8$) and class $1$ (green, $N_1=4$). The hypergraph has $E=7$ hyperedges. In Fig.~\ref{fig:example-RWHS}\textbf{(b)}, the meet-wise RWHS (Eq.~\eqref{eq:meet-RWHS}) is computed from the fraction of nodes that have the same class $c=0$ as the start node $i_0=1$, yielding $a^\mathrm{mRWHS}_{3,2}=\frac{1}{2}$. The jump-wise RWHS (Eq.~\eqref{eq:jump-RWHS}) is computed as the fraction of subsequent nodes that have the same label as each other, resulting in $a^\mathrm{jRWHS}_{3,2}=\frac{1}{4}$.

\subsubsection{Further Considerations}

Similar to MP homophily (Section~\ref{sec:message-passing-homophily}), RWHS incorporates dynamic aspects, here a stochastic process, into the measurement of homophily. This makes the measure distinct from purely structural assessments.

FRC emphasize that different definitions of random walks on hypergraphs exist and can be employed to compute both meet-wise and jump-wise RWHS. On the one hand, this opens up a vast design space for new measures~\cite{zhou2006_randomwalk, chitra2019_randomwalkshypergraphs,carletti2020_randomwalk}. On the other hand, increased caution is necessary as the chosen random walk influences the interpretation of the measure. Beyond the choice of random walk, the random walk length and the number of realizations per start node need to be determined. FRC provide little guidance on how to choose these parameters.

Finally, it is worth considering that meet-wise and jump-wise RWHS are computationally more expensive than many other homophily measures, as several realizations of random walks per node need to be simulated and the null model is not specified in closed form but through randomization of the data. Moreover, this means that hypergraph data needs to be available in a format that allows for simulations of random walks as well as randomization, not just through summary statistics.

\subsection{Hypergraph Assortativity}\label{sec:hypergraph-assortativity}

In the case of graphs, Newman's assortativity~\cite{newman2003_mixingpatterns} is among the most popular measures of mixing patterns. For continuous node attributes, it is defined as the Pearson correlation between attributes of nodes found at the two endpoints of an edge. For nominal node attributes, it measures the normalized excess agreement between nodes found at the two endpoints of an edge relative to a stub-mixing baseline, i.e., it is Cohen's $\kappa$ for edge endpoint class agreement~\cite{bojanowski2014_measuringsegregation}. 

As Newman's assortativity quantifies pairwise association, its generalization to hypergraphs is thus not immediate. However, several measures referred to as ``hypergraph assortativity'' exist.

In this section, we introduce these measures, provide an example calculation, and comment on important practical considerations and opportunities for further work.

\subsubsection{Definition}

Here, we define \textit{Chodrow's hypergraph assortativity} $\rho^{\mathrm{C}}$~\cite{chodrow2020_configurationmodel} and comment on Landry and Restrepo's \textit{dynamical hypergraph assortativity}~\cite{landry2022_hypergraphassortativity}. Both measures were originally introduced with the goal of quantifying degree correlations. While Chodrow's approach can easily be extended to deal with arbitrary continuous node attributes $x_i \in \mathcal{X}$, more work would be needed to extend dynamical hypergraph assortativity beyond degrees as it is intimately related to the impact on structure on contagion dynamics.

Chodrow defines an entire family of assortativity measures based on a potentially random choice function $\psi : \mathcal{E}\to \mathcal{V}\times \mathcal{V}$, which maps a hyperedge $e$ of arbitrary size to a tuple of distinct nodes, and a ranking function $r:\mathcal{V}\to\{1,\dots,N\}$ that associates each node with a rank based on its attribute value $x_i$. An example of a deterministic choice function is selecting the node with highest and lowest node attribute from a hyperedge, while sampling two nodes without replacement from the hyperedge is an example of a random choice function. Using the notation $\varphi_u(e)=(r\circ \psi_u)(e)$ for $u\in\{1,2\}$ and $e \in \mathcal{E}$, Chodrow's assortativity $\rho^\mathrm{C}$ is defined as the empirical Pearson correlation
\begin{equation}\label{eq:chodrow-assortativity}
    \rho^\mathrm{C}
    =
    \frac{
        \langle \varphi_1 \varphi_2\rangle_\mathcal{E} -\langle \varphi_1\rangle_\mathcal{E} \langle \varphi_2\rangle_\mathcal{E}
    }{
        \sqrt{
            (\langle \varphi_1 \varphi_1\rangle_\mathcal{E} -\langle \varphi_1\rangle_\mathcal{E} \langle \varphi_1\rangle_\mathcal{E})
            (\langle \varphi_2 \varphi_2\rangle_\mathcal{E} -\langle \varphi_2\rangle_\mathcal{E} \langle \varphi_2\rangle_\mathcal{E})
            }
    }
    ,
\end{equation}
with the empirical average over all hyperedges
\begin{equation}
    \langle \varphi_u \rangle_\mathcal{E}
    =
    \frac{1}{E}
    \sum_{e \in \mathcal{E}}
    \varphi_u(e)
    .
\end{equation}
This means Chodrow's assortativity is an empirical Spearman-type correlation between the attributes of two nodes selected from a random hyperedge by a choice function. Positive values of the measure indicate homophily, while negative values indicate heterophily.

An alternative assortativity coefficient is introduced by Landry \& Restrepo, who refer to it as dynamical hypergraph assortativity. The quantity is named for its influence on contagion dynamics on higher-order networks with degree correlations. As degree is not an arbitrary node attribute but tied to the hypergraph's structure, replacing degree with other node attributes does not preserve the measure's interpretation in the context of contagion dynamics. Extending dynamical assortativity to node attributes beyond degree, e.g., using an appropriate hypergraph model in which hypergraph structure depends on node attributes, is a promising direction for future work but beyond the scope of this chapter.

\subsubsection{Example}

\begin{figure}
    \centering
    \includegraphics[width=0.99\linewidth]{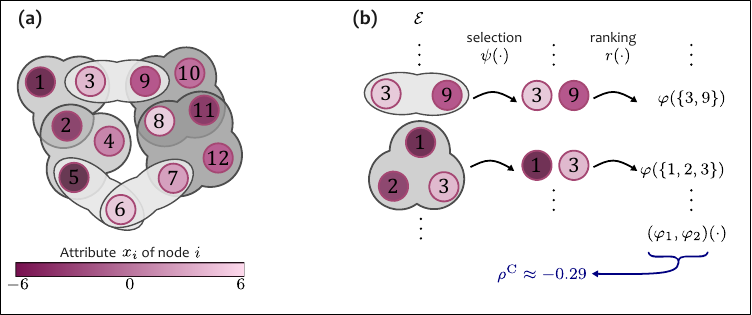}
    \caption{\textbf{Hypergraph Assortativity:} \textbf{(a)} An example hypergraph $\mathcal{H}$ with $N=12$ nodes. Each node $i$ is associated with a scalar attribute $x_i\in[-6,6]$ (color coded). \textbf{(b)} Chodrow's assortativity (Eq.~\eqref{eq:chodrow-assortativity}) is computed by applying the choice function $\psi$, here selecting the minimum and maximum attribute value, to each hyperedge, ranking the selected nodes according to their attribute using the rank function $r$, and computing the empirical correlation between ranks of nodes selected from the same hyperedge. This measure indicates disassortativity or heterophily, $\rho^\mathrm{C}\approx -0.29$.}
    \label{fig:example-assortativity}
\end{figure}

In Fig.~\ref{fig:example-assortativity}, we illustrate how to calculate hypergraph assortativity using Chodrow's (Eq.~\eqref{eq:chodrow-assortativity}) definition. We consider an example hypergraph $\mathcal{H}$ (Fig.~\ref{fig:example-assortativity}\textbf{(a)}) with a real node attribute $x\in [-6, 6]$. Chodrow's assortativity is calculated by selecting two nodes from each hyperedge (Fig.~\ref{fig:example-assortativity}\textbf{(b)}). In this example, the choice function selects the nodes with largest and smallest attribute value. Then, the rank function assigns each selected node its rank according to the attribute value. Finally, the empirical correlation coefficient is computed. In our example, this results in a value of $\rho^\mathrm{C}\approx-0.29$, meaning that, on average, higher maximum attribute values within a hyperedge are associated with lower minimum values.

\subsubsection{Further Considerations}

At least two definitions for hypergraph assortativity exist in the literature, raising the question of whether a unique best measure for this property exists. However, only one of these measures, Chodrow's hypergraph assortativity, is directly applicable to attributes other than node degree. Extending Landry and Restrepo's approach to attributes other than degree is an important open problem.

Designing measures for higher-order networks based on pairwise correlation coefficients faces the challenge of creating observation pairs. Chodrow's assortativity solves this using the choice function $\psi$, which is thus key to understanding what types of structure this measure can capture. In contrast, the use of higher-order correlations, in the sense of higher raw moments or correlators, as an alternative to pairwise correlation coefficients remains underexplored. 

Chodrow's assortativity is a hypergraph-level quantity. Exploring more fine-grained versions of the measure, e.g., a hyperedge size-dependent formulation, is an interesting opportunity for future work. In this context, related work by Chodrow and Mellor discusses the possibility of stratifying averages by the role of nodes (a concept that encompasses node classes as a special case)~\cite{chodrow2020_annotatedhypergraphs}. 

While Chodrow's assortativity closely mimics Newman's assortativity for the case of size-$2$ hyperedges, apart from using a Spearman-type correlation instead of Pearson correlation, it does not directly provide an analogue of Newman's nominal assortativity. However, such an extension appears relatively straightforward by exchanging correlations with chance-corrected agreement scores.

It is also worth noting that while Newman's assortativity coefficients are widely used, they have also been criticized from various perspectives, e.g., nominal assortativity might not be adequate when class imbalance exists~\cite{karimi2023_inadequacyassortativity} or on networks with heavy-tailed degree distributions in which node degrees and attributes are correlated~\cite{litvak2013_uncoveringdisassortativitylarge, cinelli2020_networkconstraints}. Exploring these limitations in the context of hypergraphs remains an open problem.

\subsection{Hypergraph Modularity}\label{sec:hypergrah-modularity}

While the most prominent application of modularity is as an objective function in community detection~\cite{newman2004_modularity, fortunato2010_communitydetection, chakraborty2017_communitydetection, peixoto2023_descriptivevsinferential}, it can also be used to assess how strongly assortative a fixed partition of nodes into classes is or to compare partitions with respect to different node classes or attributes. 

Generalizing modularity to hypergraphs faces similar problems as other measures of mixing patterns: The dichotomy between within-class and between-class edges breaks down in hyperedges of size greater than two. The two main proposals for hypergraph modularity thus define entire families of measures, each resolving this issue differently~\cite{kaminski2024_modularity-2, kaminski2019_modularity-1, chodrow2021_generativehypergraphclustering}.

Here, we introduce the hypergraph modularity of Kami\'nski, Misiorek, Pralat, and Th\'eberge (KMPT)~\cite{kaminski2024_modularity-2, kaminski2019_modularity-1}, and illustrate its use through an example computation. For an alternative definition based on maximum (pseudo-)likelihood estimation of the parameters of the degree-corrected hypergraph stochastic block model (DC-HSBM), we refer to the work of Chodrow, Veldt, and Benson~\cite{chodrow2021_generativehypergraphclustering}.

\subsubsection{Definition}

On graphs, modularity compares the number of within-class edges to their expected number under a degree-corrected baseline~\cite{newman2004_modularity}. Hypergraph modularity mirrors this pattern by comparing the number of  $(c, \alpha, t)$-hyperedges to their expected count $\bar{E}_{\alpha, t}^{(c)}$ under a null model. The null model considers $E_\alpha$ hyperedges of size $\alpha$ and assigns their members independently at random to class $c \in \mathcal{C}$ with the probability $\pi_c$ that a node is assigned to class $c$ determined by the degrees of nodes in class $c$,
\begin{equation}
    \pi_c =         \frac{
            \sum_{i \in \mathcal{V}_c} d(i)
        }{
            \sum_{i \in \mathcal{V}} d(i)
        }.
\end{equation} 
The \textit{hypergraph modularity} $Q_{\tau}^{\mathrm{KMPT}}$ aggregates the differences between observed and expected counts over $(c, \alpha, t)$-hyperedges in which class $c$ constitutes a majority,
\begin{equation}\label{eq:kmpt-modularity}
\begin{split}
    Q^\mathrm{KMPT}_{\tau}
    &=
    \sum_{\alpha = 2}^{\alpha_\mathrm{max}}
    \sum_{t = \lfloor \frac{\alpha}{2} \rfloor +1}^\alpha
    \left(
        \frac{t}{\alpha}
    \right)^{\tau}
    Q_{\alpha,t}^\mathrm{KMPT}\\
    &=
    \sum_{\alpha = 2}^{\alpha_\mathrm{max}}
    \sum_{t = \lfloor \frac{\alpha}{2} \rfloor +1}^\alpha
    \left(
        \frac{t}{\alpha}
    \right)^{\tau}
    \frac{1}{E}
    \sum_{c \in \mathcal{C}}
    \left(
        E^{(c)}_{\alpha,t}
        -
        E_\alpha
        \binom{\alpha}{t}
        \pi_c^t
        (1 - \pi_c)^{\alpha - t}
    \right),
\end{split}
\end{equation}
where $Q_{\alpha,t}^\mathrm{KMPT}$ are the contributions of hyperedges of size $\alpha$ and type $t$ with respect to any class $c$ to the total modularity $Q_\tau^\mathrm{KMPT}$, and $\tau \in [0, \infty]$ is a hyperparameter that controls the weighting of these contributions. Intuitively, $\tau$ determines which hyperedge types are declared within-class: For $\tau = \infty$ only homogeneous hyperedges are dubbed within-class. For $\tau = 0$ all hyperedges, in which a given class has a majority, are considered within-class hyperedges and taken into account with equal weight.

\subsubsection{Example}

\begin{figure}
    \centering
    \includegraphics[width=0.99\linewidth]{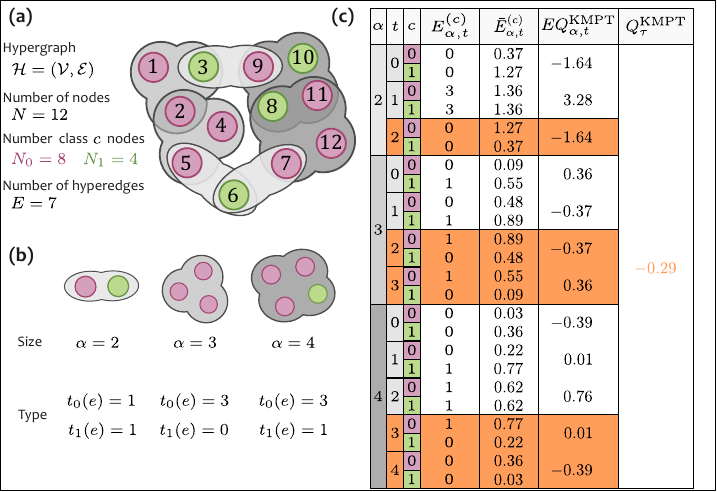}
    \caption{\textbf{Hypergraph Modularity:} \textbf{(a)} A hypergraph $\mathcal{H}$ with a partition of its $N=12$ nodes into two classes, class $0$ (purple) with $N_0=8$ and class $1$ (green) with $N_1=4$ nodes. The number of hyperedges is $E=7$. \textbf{(b)} In order to compute hypergraph modularity, the type $t_c(e)$ of each hyperedge $e \in \mathcal{E}$ with respect to each class $c \in \{0, 1\}$, i.e., the number of class $c$ nodes in it, needs to be determined. The type of a hyperedge determines whether and how strongly, it is counted as a within- or between-class hyperedge. \textbf{(c)} Hypergraph modularity $Q_{\tau}^\mathrm{KMPT}$ is computed from the number of hyperedges $E_{\alpha, t}^\mathrm{(c)}$ of size $\alpha$ and type $t$ with respect to class $c$ and the expected number of such edges $\bar{E}_{\alpha,t}^{(c)}$. Only hyperedge counts $E_{\alpha, t}^\mathrm{(c)}$ in which $c$ constitutes a majority (orange rows) enter the modularity calculation. The overall modularity $Q^{\mathrm{KMPT}}_\tau$ is computed as a (weighted) average of the contributions from different hyperedge sizes $\alpha$ and types $t$ according to Eq.~\eqref{eq:kmpt-modularity}. Weighting all contributions equally, i.e., setting $\tau=0$, yields $Q=-0.29$. This means the hypergraph has fewer within-class hyperedges than expected under the null model.}
    \label{fig:example-modularity}
\end{figure}

To illustrate how hypergraph modularity is computed, we consider a small example hypergraph $\mathcal{H}$ with $N=12$ nodes, $N_0=8$ of which come from class $0$ and $N_1=4$ of which come from class $1$. The total number of hyperedges is $E=7$ (Fig.~\ref{fig:example-modularity}\textbf{(a)}). To compute the hypergraph modularity, the type $t_c(e)$ of each hyperedge $e$, i.e., the number of class $c$ nodes in $e$ (Eq.~\eqref{eq:type}), needs to be determined (Fig.~\ref{fig:example-modularity}\textbf{(b)}). Counting the number $E_{\alpha , t}^{(c)}$ of hyperedges of each size $\alpha \in \{2,3,4\}$ and type $t\in \{0, \dots, \alpha\}$ with respect to each class $c \in \{0,1\}$ and comparing them to their expected number $\bar{E}_{\alpha ,t}^{(c)}$, yields the size- and type-dependent contribution to the modularity, $Q_{\alpha, t}^\mathrm{KMPT}$ (Fig.~\ref{fig:example-modularity}\textbf{(c)}). Only contributions from hyperedge counts $E_{\alpha,t}^{(c)}$ for which class $c$ is the majority (orange rows) enter the hypergraph modularity calculation. In this example, we view all hyperedges in which a given class constitutes a majority as within-class hyperedges of this class and weight their contributions to the total hypergraph modularity equally (i.e., we set $\tau =0$ in Eq.~\eqref{eq:kmpt-modularity}). This yields an overall modularity of $Q_{0}^{\mathrm{KMPT}} = - 0.29$, indicating that within-class hyperedges are less abundant than expected under the null model. According to KMPT's hypergraph modularity with hyperparameter $\tau =0$, the hypergraph is thus heterophilous.

\subsubsection{Further Considerations}

Modularity is a commonly used yet controversial tool in network science: In particular, its use as an objective function in community detection has been heavily criticized, e.g., due to an inherent resolution limit for identifying small communities~\cite{fortunato2007_resolutionlimit} and due to the risk of overfitting to noise in the network structure~\cite{peixoto2023_descriptivevsinferential}. Only some of these criticisms are also applicable when using modularity for measurement instead of optimization. For example, while it is still worth acknowledging that hypergraph data is noisy, there is no immediate overfitting issue. 

We emphasize that raw hypergraph modularity $Q^\mathrm{KMPT}_\tau$ should only be used to compare different partitions with a comparable number of classes on the same hypergraph. To enable comparison across hypergraphs or partitions with a different number of classes, one would need to create a properly normalized score. In the case of graphs, this is exactly what Newman's nominal assortativity does~\cite{newman2003_mixingpatterns, newman2004_modularity}. It rescales modularity by its maximum possible value, over stub-matching configurations rather than label assignments. Determining an appropriate and easily comparable normalization is an open problem in the case of hypergraph modularity. 

While (hypergraph) modularity as such might not be a convenient tool for assessing homophily, the rich literature dealing with this measure in the pairwise and higher-order cases can serve as inspiration for the design of hypergraph homophily measures.

\subsection{Pairwise Homophily of the Clique Projection}\label{sec:clique-projection-homophily}

As homophily measures for graphs are abundant (see, e.g.,~\cite{bojanowski2014_measuringsegregation, saxena2025_homophilytutorial}), it might thus seem attractive to transform a hypergraph via clique projection into a graph and leverage existing measures (an approach used, e.g., in~\cite{wang2022_equivarianthypergraphdiffusion}).

Here, we explain what this approach entails, illustrate it through an example, and provide guidance on when---if at all---this approach is appropriate.

\subsubsection{Definition}

In order to use pairwise homophily measures in the context of higher-order interactions, two steps are necessary: First, the hypergraph needs to be transformed into a graph. Second, a homophily measure for graphs needs to be selected and applied to the resulting graph.

To transform a hypergraph $\mathcal{H}=(\mathcal{V},\mathcal{E})$ into a graph $\mathcal{H}^{\downarrow}=(\mathcal{V},\mathcal{E}')$, we rely on the \textit{clique projection} (sometimes also called clique expansion), i.e., $\mathcal{H}$ and $\mathcal{H}^{\downarrow}$ have the same node set $\mathcal{V}$, and two distinct nodes $i$ and $j$ are connected with an undirected edge, i.e., $(i,j)\in\mathcal{E}'$ if and only if they co-occur in at least one hyperedge $e \in \mathcal{E}$. Note that $\mathcal{H}^{\downarrow}$ defined in this way is a simple graph, i.e., contains neither self-loops nor multi-edges.
To compute homophily, any homophily measure developed for graphs could now be employed, e.g., the homophily index~\cite{coleman1958_homophily, currarini2009_economicmodelfriendship, altenburger2018_monophily}, nominal assortativity~\cite{newman2003_mixingpatterns}, or BA-homophily~\cite{lee2019_perceptionbias, karimi2018_homophilyranking, lee2024_comparison}.
\subsubsection{Example}

\begin{figure}
    \centering
    \includegraphics[width=0.99\linewidth]{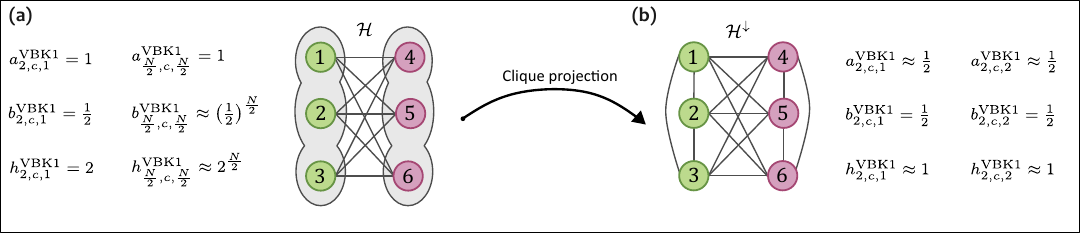}
    \caption{\textbf{Clique Projection:}
    \textbf{(a)} An example hypergraph $\mathcal{H}$ with $N=6$ nodes equally divided into two classes $0$ (purple) and $1$ (green), $N_0=N_1=\frac{N}{2}=3$. The sub-hypergraph $\mathcal{H}^{(2)}$ of size-$2$ hyperedges is the complete bipartite graph, and the sub-hypergraph $\mathcal{H}^{\left(\frac{N}{2}\right)}$ of size-$\frac{N}{2}$ hyperedges consists of two hyperedges each encompassing the nodes of exactly one class. Computing the standard ratio scores of hypergraph homophily (Eq.~\eqref{eq:vbk-standard-ratio-score}) indicates strong heterophily at hyperedge size $2$ and strong homophily at hyperedge size $\frac{N}{2}$.
    \textbf{(b)} Analyzing homophily in the clique projection $\mathcal{H}^\downarrow$ of $\mathcal{H}$ yields a very different result: All standard affinity scores approach their baseline values as $N \to \infty$, indicating a neutral hypergraph.}
    \label{fig:example-clique-projection}
\end{figure}

Here, we illustrate through an admittedly extreme example how the clique projection can erase information about the homophily pattern of a hypergraph. Consider a hypergraph $\mathcal{H}_N$ with $N$ nodes. We assume that $N$ is even and that the first $N_0=\frac{N}{2}$ nodes belong to class $0$, and the second $N_1=\frac{N}{2}$ nodes belong to class $1$. These nodes are connected through hyperedges of size $\alpha = 2$ and $\alpha = \frac{N}{2}$ as follows: The sub-hypergraph of size $\alpha = 2$ hyperedges, $\mathcal{H}^{(2)}$, is the complete bipartite graph, i.e., all possible between-class hyperedges are present but no within-class hyperedges exist. The sub-hypergraph of $\alpha = \frac{N}{2}$ hyperedges, $\mathcal{H}^{\left(\frac{N}{2}\right)}$, consists of $E_\frac{N}{2}=2$ hyperedges. One hyperedge encompasses all nodes in class $0$, and the other one encompasses all nodes in class $1$. Figure~\ref{fig:example-clique-projection}\textbf{(a)} illustrates this hypergraph in the case of $N=6$ nodes.

VBK's standard ratio scores (Eq.~\eqref{eq:vbk-standard-ratio-score}) for size-$2$ hyperedges are equal for both classes, and indicate strong heterophily, $h^{\mathrm{VBK}1}_{c, 2, 1}=2$, which can be seen from explicit computation of the standard affinity scores (Eq.~\eqref{eq:vbk-standard-affinity-scores}) and asymptotic standard baseline scores (Eq.~\eqref{eq:vbk-standard-baseline-scores}),
\begin{equation}
\begin{split}
    a^{\mathrm{VBK}1}_{c, 2, 1}(\mathcal{H})
    &=
    \frac{N_c}{N_c}
    =
    1 \\
    b^{\mathrm{VBK}1}_{c, 2, 1}(\mathcal{H})
    &=
    \binom{2 - 1}{1 - 1} \left(\frac{1}{2}\right)^{1-1} \left(1 - \frac{1}{2} \right)^{2 - 1}
    =
    \frac{1}{2},
\end{split}
\end{equation}
where we explicitly indicated the hypergraph $\mathcal{H}$ for which the scores are computed. In contrast, the standard ratio scores for size-$\frac{N}{2}$ hyperedges indicate strong (even diverging) homophily $h^{\mathrm{VBK}1}_{c, \frac{N}{2},\frac{N}{2}}= 2^{\frac
{N}{2}-1}$, based on the standard affinity scores and asymptotic standard baseline scores
\begin{equation}
\begin{split}
    a^{\mathrm{VBK}1}_{c, \frac{N}{2}, \frac{N}{2}}(\mathcal{H})
    &=
    \frac{2N_c}{2N_c}
    =
    1 \\
    b_{c, \frac{N}{2}, \frac{N}{2}}^{\mathrm{VBK}1}(\mathcal{H})
    &=
    \binom{N_c - 1}{N_c - 1} \left(\frac{1}{2}\right)^{\frac{N}{2} - 1} \left(1 - \frac{1}{2} \right)^{0}
    =\left(\frac{1}{2}\right)^{\frac{N}{2}-1}.
\end{split}
\end{equation}
However, after clique projection, illustrated in Fig.~\ref{fig:example-clique-projection}\textbf{(b)} for the case of $N=6$ nodes, we obtain a very different result. We again use VBK's standard affinity scores as they are equal to the graph homophily index if the hyperedge size is $\alpha = 2$. The standard ratio scores indicate an asymptotically neutral hypergraph, $h_{c, 2,1}^{\mathrm{VBK}1}, h_{c,2,2}^{\mathrm{VBK}1}\to1$ with respect to either class. This can be seen through explicit computation of the standard affinity scores and asymptotic standard baseline scores for size-$2$ hyperedges,
\begin{equation}
\begin{split}
    a^{\mathrm{VBK}1}_{c, 2, 2}(\mathcal{H}^{\downarrow})
    &=
    \frac{
            2 \binom{N_c}{2}
        }{
            2 \binom{N_c}{2} + N_c^2
    }
    =
    \frac{
        (N_c - 1)
    }{
        (N_c - 1) + N_c
    }
    \overset{
        N_c \to \infty
    }{
        \longrightarrow
    }
    \frac{1}{2} \\
    b_{c, 2, 2}^{\mathrm{VBK}1}(\mathcal{H}^{\downarrow})
    &=
    \binom{2 - 1}{2 - 1} \left(\frac{1}{2}\right)^{2-1} \left(1 - \frac{1}{2} \right)^{2-2}
    =
    \frac{1}{2} \\
    a^{\mathrm{VBK}1}_{c, 2, 1}(\mathcal{H}^{\downarrow})
    &=
    \frac{
            N_c^2
        }{
            2 \binom{N_c}{2} + N_c^2
    }
    =
    \frac{
        N_c
    }{
        (N_c - 1) + N_c
    }
    \overset{
        N_c \to \infty
    }{
        \longrightarrow
    }
    \frac{1}{2}\\
    b^{\mathrm{VBK}1}_{c, 2, 1}(\mathcal{H}^{\downarrow})
    &=
    \binom{2 - 1}{1 - 1} \left(\frac{1}{2}\right)^{1-1} \left(1 - \frac{1}{2} \right)^{2 - 1}
    =
    \frac{1}{2}.
\end{split}
\end{equation}

This means depending on whether we conduct our analysis using hypergraphs or graphs obtained by clique projection, we reach vastly different conclusions: The hypergraph perspective reveals that vastly different connection preferences exist based on the size of the interactions. The graph perspective shows us that nodes are, overall, in contact with nodes from either class in approximately equal proportions. 

\subsubsection{Further Considerations}

Employing the clique projection in combination with homophily measures for graphs is a double-edged sword. On the one hand, it gives access to the wide variety of measures developed for the pairwise case. These measures are far more studied, so their respective merits and drawbacks are well characterized~\cite{bojanowski2014_measuringsegregation}. On the other hand, it erases information about the hyperedge size dependence of homophily.

Overall, we advocate for using hypergraph-native homophily measures when analyzing hypergraph data. Choosing a hypergraph representation for a system means committing to the assumption that group interactions, i.e.,  hyperedges are meaningful beyond the sum of all pairwise interactions of the nodes within them. Computing homophily via clique projection and pairwise homophily measures runs counter to this assumption and raises the question of why the data wasn't represented as a graph in the first place. Furthermore, it is questionable whether the underlying null models of many pairwise homophily measures are suitable for graphs resulting from clique projection, as treating edges resulting from the expansion of the same hyperedge as independent observations seems poorly justified.

Finally, we point out that clique projection is not the only strategy for employing pairwise homophily measures directly on hypergraphs. An alternative approach is discussed by FRC~\cite{failla2025_RHWS}.


\section{Hypergraph Models for Studying Mixing Patterns}\label{sec:hypergraph-models}

Hypergraph models are indispensable for the study of homophily in the higher-order setting~\cite{cimini2019_statisticalphysics, peel2022_statisticalinference}. A hypergraph model is a probability distribution $P(\mathcal{H})$ over some set of hypergraphs, called the \textit{sample space}. Hypergraph models have multiple purposes: First, they specify what we mean by ``expected at random'' and can thus serve as null models for defining homophily measures. Second, they can be fit to data, thereby offering insight into the mixing patterns of real-world systems either through interpretable parameters or model comparison. In fact, some of the measures introduced in Section~\ref{sec:homophily-measures} are intimately related to the parameters of different hypergraph models. Finally, if a model admits a (computationally efficient) sampling procedure, new hypergraphs can be generated from the model. These can be used in computational studies, e.g., agent-based models of opinion dynamics. We emphasize that it is important not to confuse a hypergraph model with the algorithm used to sample from the model.

This section provides an overview of hypergraph models that are useful for studying higher-order mixing patterns. We do not aim to comprehensively describe all hypergraph models that could potentially be used to this end but rather provide examples for four main classes: null models that do not account for homophily, hypergraph stochastic block models in which nodes are assigned to discrete classes, geometric hypergraph models in which nodes are characterized by continuous latent positions, and models that implicitly specify probability distributions through algorithmic rules and often admit more efficient hypergraph generation.

\subsection{Hypergraph Null Models}\label{sec:null models}

Studying homophily in hypergraphs not only requires models that are able to generate hypergraphs with flexible class-mixing patterns, but also null models of hypergraphs to understand what properties are explained by simple constraints. Such models are usually formulated as maximum-entropy models~\cite{cimini2019_statisticalphysics, jaynes1982_maxent, park2004_statisticalmechanicsnetworks}. A standard distinction among these models is between models that fix certain hypergraph properties in expectation (canonical models) and those that fix such properties exactly (micro-canonical models).

For graphs, the arguably simplest null models are the canonical and microcanonical \ER{} models, which fix the number of edges either exactly or in expectation~\cite{gilbert1959_randomgraphs, erdos2011_evolutionrandomgraphs}. In the hypergraph setting, one needs to fix the number of hyperedges $E_\alpha$ of each hyperedge size $\alpha$ either exactly or in expectation~\cite{schmidtpruzan1985_ERhypergraphs}. These \textit{\ER{}} \textit{hypergraphs} are models of potentially non-uniform, simple hypergraphs. In contrast, the \textit{random hypergraph model} (RHM)~\cite{saracco2025_randomhypergraphmodel} considers a hypergraph with $N$ nodes and $E$ hyperedges in a bipartite graph representation and constrains the number of nonzero entries in the hypergraph's incidence matrix. This leads to a model of hypergraphs with parallel and potentially non-simple hyperedges.

As hypergraphs arising in many real-world modeling applications exhibit heavy-tailed degree distributions, constraints on a node's (expected) degree can be of interest~\cite{lee2025_surveyhypergraphmining, do2020_StructuralPatternsGenerative}. In the case of graphs, the relevant models are called configuration models~\cite{park2004_statisticalmechanicsnetworks, chung2002_averagedistancesrandom, molloy1995_criticalpointrandom, fosdick2018_configuringrandomgraph}. The analogue of the canonical configuration model is the $\beta$-\textit{model for hypergraphs}, which treats either a node $i$'s degree $d(i)$ or its $\alpha$-degree $d_\alpha(i)$ at each hyperedge size $\alpha$ as fixed in expectation~\cite{stasi2014_betamodel}. In the former case, the expected number of hyperedges of each size is a property of the model; in the latter case, it is fixed implicitly by specifying all $d_\alpha(i)$. The sample space of this model is the set of potentially non-uniform, simple hypergraphs with $N$ nodes. An alternative canonical configuration model for hypergraphs was introduced by Saracco et al.~\cite{saracco2025_randomhypergraphmodel}. This model is formally identical to a specific configuration model for bipartite graphs~\cite{saracco2015_randomizingbipartitenetworks}, admitting only simple but potentially parallel hyperedges. Both the $\beta$-model and Saracco et al.'s model permit Chung-Lu-like approximations in sufficiently sparse regimes~\cite{chung2002_averagedistancesrandom}.

Microcanonical configuration models are uniform distributions on all hypergraphs in a sample space specified through hard constraints on node degrees and potentially hyperedge sizes. As uniform sampling from such spaces by generating hypergraphs from scratch is challenging, most configuration models are sampled through Markov Chain Monte Carlo (MCMC) algorithms that rewire a given hypergraph. As in the case of graphs, specifying the sample space precisely is paramount as it determines not only the exact algorithm but also the feasibility of MCMC sampling in the first place~\cite{fosdick2018_configuringrandomgraph}. Configuration models on stub- and node-labeled hypergraphs with simple, potentially parallel hyperedges and an MCMC algorithm for sampling from these models were introduced by Chodrow~\cite{chodrow2020_configurationmodel}. An algorithmic alternative was recently formulated under the name of \textit{VaLUH}~\cite{abuissa2026_valuh}. More general sample spaces are discussed by Kraakman and Stegehuis~\cite{kraakman2025_hypercurveball, kraakman2026_uniformsampling} as well as Preti et al.~\cite{preti2024_higherordernullmodels} in the context of directed hypergraphs, and for hypergraphs with hyperedge-dependent node roles by Chodrow and Mellor~\cite{chodrow2020_annotatedhypergraphs}. Even though they are not hypergraph models per se, colored configuration models for graphs could play an important role in developing hypergraph configuration models further~\cite{preti2025_polaris, pellegrina2026_CCM}.

The $dK$-series for graphs builds a hierarchy of increasingly strong hard constraints on subgraph counts and node degrees~\cite{mahadevan2006_systematictopologyanalysis, orsini2015_quantifyingrandomness}. Nakajima, Shudo, and Masuda extend these ideas to hypergraphs~\cite{nakajima2022_randomizinghypergraphs} using a hypergraph's bipartite representation.

\subsection{Hypergraph Stochastic Block Models \& Discrete Node Attributes}\label{sec:hypergraph-sbm}

Hypergraph stochastic block models permit flexible mixing patterns among nodes belonging to discrete classes. We highlight a few models that are conceptually interesting for the study of homophily, rather than comprehensively surveying the literature, thus neglecting theoretical results on community recovery~\cite{angelini2015_spectraldetectionsparse, stephan2022_sparserandomhypergraphs, gu2023_weakrecoverythreshold} or hypergraph limits~\cite{balasubramanian2021_nonparametricghigherorder}. In all models below, the existence (or multiplicity) of each hyperedge is an independent random variable (conditioned on model parameters) making the probability of a hypergraph easily accessible in closed form. However, the large number of hyperedges that could possibly exist can make na\"ive sampling from these models challenging for large number of nodes $N$ and even moderate hyperedge sizes $\alpha$. Even if generating hypergraphs can pose a challenge, inferring model parameters such as the class membership of nodes can be possible, and hypergraph stochastic block models are frequently used (and arguably developed) with this goal in mind. While hypergraph stochastic block models are undeniably popular, we acknowledge that they are not the only statistically well-motivated models for hypergraphs with node-wise discrete latent variables (see, e.g., the \textit{ELCA model} in~\cite{ng2022_modelbasedclustering}).

Ghoshdastidar and Dukkipati~\cite{ghoshdastidar2014_HSBM-1, ghoshdastidar2017_HSBM-2} are widely credited with introducing the basic \textit{hypergraph stochastic block model} (HSBM). The model describes non-uniform hypergraphs with $N$ nodes each assigned to one of $C$ classes, excluding parallel or non-simple hyperedges. The probability $p_{i_1,\dots, i_\alpha}^{(\alpha)}$ that a size-$\alpha$ hyperedge among nodes $i_1 < i_2 < \dots < i_\alpha$ exists depends only on the class membership of the nodes through a symmetric order-$\alpha$ tensor $\pi^{(\alpha)}$,
\begin{equation}
    p_{i_1, \dots, i_\alpha}^{(\alpha)}
    =
    \pi_{c_{i_1}, \dots, c_{i_\alpha}}^{(\alpha)}.
\end{equation}
As the tensors $\pi^{(\alpha)}$ are symmetric, only the abundances of a given class within the hyperedge---not its ``position''---matters. In analogy to the graph case, this model is a maximum-entropy model with a given expected number of hyperedges with specific size and class composition. As each tensor $\pi^{(\alpha)}$ has $\binom{C + \alpha -1}{C - 1}$ independent entries, the model can be challenging to specify in practice, prompting VBK to introduce the \textit{cardinality-based hypergraph stochastic block model} (CB-HSBM) for the case of $C=2$ classes~\cite{veldt2023_hypergraphhomophily}. The CB-HSBM is the natural model for hypergraphs with a given number of $(c, \alpha, t)$-hyperedges with respect to an arbitrarily chosen reference class $c$, and allows us to explore statistical interpretations of VBK's hypergraph homophily (Section~\ref{sec:hypergraph-homophily}).

Investigating how degree heterogeneity and class mixing jointly shape higher-order networks requires defining a degree-corrected version of the HSBM. The \textit{degree-corrected hypergraph stochastic block model} (DC-HSBM)~\cite{chodrow2021_generativehypergraphclustering} is a model of non-uniform hypergraphs with $N$ nodes, each belonging to one of $C$ classes, that allows for non-simple and parallel hyperedges. The multiplicity of a size-$\alpha$ hyperedge among nodes $i_1\leq i_2 \leq \dots \leq i_\alpha$ is Poisson distributed with a rate $\lambda_{i_1,\dots,i_\alpha}^{(\alpha)}$ depending on the class membership of nodes through a symmetric tensor $\Omega^{(\alpha)}$ and an expected-degree parameter $\theta_i$ for each member node $i$,
\begin{equation}
    \lambda_{i_1,\dots,i_\alpha}^{(\alpha)}
    =
    \eta_{i_1,\dots,i_\alpha}^{(\alpha)}
    \left(
        \prod_{i \in \{\!\{i_1, \dots, i_\alpha \}\!\}}
        \theta_i
    \right)
    \Omega_{c_{i_1},\dots, c_{i_\alpha}}^{(\alpha)}
    ,
\end{equation}
where $\eta^{(\alpha)}$ is the number of ways to form distinct ordered tuples from the entries of the multiset $\{\!\{i_1,\dots, i_\alpha\}\!\}$. For both generative modeling and inference, it is interesting to impose further constraints on $\Omega^{(\alpha)}$ to reduce the number of independent parameters, e.g., $\Omega^{(\alpha)}$ could be constrained to depend on relative class imbalance instead of the particular classes in a hyperedge. If non-simple and parallel hyperedges are undesirable for a particular application, the degree-corrected block model for hypergraphs (hDCBM)~\cite{ke2020_hDCBM} offers an alternative to the DC-HSBM.

Many homophily measures assume that nodes can only belong to a single class. Mixed-membership stochastic block models move beyond this assumption and allow node $i$ to act as a member of class $c$ with probability $\zeta_{ic}$. In the \textit{mixed-membership hypergraph stochastic block model} (HyGMMSBM)~\cite{salespardo2023_HyGMMSBM}, the probability $p_{i_1,\dots, i_\alpha}^{(\alpha)}$ that a size-$\alpha$ hyperedge among nodes $i_1 < i_2 < \dots < i_\alpha$ exists is governed by the parameters $\{\zeta_{ic}\}_{c=1}^C$ for these nodes and a symmetric order-$\alpha$ tensor $\pi^{(\alpha)}$ encoding the probability of nodes from certain classes to interact,
\begin{equation}
    p_{i_1,\dots, i_\alpha}^{(\alpha)}
    =
    \sum_{c_1,\dots,c_\alpha=1}^C
    \left(
        \prod_{u=1}^\alpha
        \zeta_{i_u c_u}
    \right)
    \pi^{(\alpha)}_{c_1,\dots,c_\alpha}
    .
\end{equation}
In this form, hyperedges are either absent or present, but several versions of this model for hypergraphs with parallel hyperedges exist. However, they often make restrictive assumptions on the class-affinity tensor in the interest of efficient inference~\cite{contisciani2022_HypergraphMT, ruggeri2023_HyMMSBM} or hypergraph generation~\cite{ruggeri2024_frameworkforgenerating}. Only recently have models been introduced that try to achieve the best of both worlds~\cite{hood2026_omnihypesmt, nakajima2025_HyperMOSBM}.

\subsection{Geometric Hypergraph Models \& Continuous Node Attributes}

Hypergraph models that derive the probability of a hyperedge from the position of its constituent nodes in a latent space naturally account for attributes that can vary gradually. One could, for example, think of continuously varying political leaning, as opposed to discrete party membership. We refer to such models as \textit{geometric hypergraph models}~\cite{boguna2021_networkgeometry}.
Our goal is to convey different ways of constructing such models, rather than providing a comprehensive overview of the existing literature. We emphasize two axes along which geometric models can differ: The first axis is the model's conditional independence assumptions. We focus on models in which either elements of the adjacency tensor or the incidence matrix are independent random variables conditioned on node (and if applicable hyperedge) positions, but other assumptions are possible (see e.g.,~\cite{yu2025_diph}). For example, models that assume the entries of the adjacency tensor are independent given node coordinates are the natural analogue of latent space models in the sense of Hoff et al.~\cite{hoff2002_latentspaceapproaches}. The second axis is the mechanism that translates latent positions into hypergraph structure, including distances and multi-linear forms. We do not include models that are restricted to fully-nested hypergraphs (reviewed in~\cite{bobrowski2022_randomsimplicialcomplexes}), as well as models for general hypergraphs directly based on them (e.g.,~\cite{turnbull2024_latentspacemodeling}).

The \textit{hypergraph latent space model} (hyper-LSM)~\cite{lyu2023_hLSM} provides an example of a model of $\alpha$-uniform hypergraphs, in which hyperedges are independent random variables conditioned on node coordinates and parameters, and latent node positions govern node existence probabilities through a nonlinear transformation of a multi-linear form: Each node $i\in \mathcal{V}$ is assigned a latent position, $x_i \in \mathbb{R}^d$, and a size-$\alpha$ hyperedge between nodes $i_1 \leq i_2 \leq \dots \leq i_\alpha$ forms with probability,
\begin{equation}
    p^{(\alpha)}_{i_1,\dots, i_\alpha}
    =
    \sigma
    \left( 
        \sum_{u_1,\dots, u_\alpha = 1}^d
        C_{u_1,\dots, u_\alpha}^{(\alpha)}
        x_{i_1, u_1} \cdots x_{i_\alpha, u_\alpha} 
    \right)
\end{equation}
where $\sigma : \mathbb{R} \to [0,1]$ is called the link function, and the symmetric order-$\alpha$ tensor $C^{(\alpha)} \in \mathbb{R}^{d\times \dots \times d}$ is called the interaction tensor. Note that non-simple hyperedges are possible, but parallel hyperedges are not. Depending on the interaction tensor, the model can accommodate flexible mixing patterns beyond homophily. The authors study the model from the perspective of inference, i.e., estimating $C^{(\alpha)}$ and $\{x_i\}_{i\in\mathcal{V}}$ from data. Sampling hypergraphs from the model can be challenging, as the time complexity of computing hyperedge probabilities is exponential in the hyperedge size, $\mathcal{O}(d^\alpha)$. This problem can be alleviated with structural assumptions on $C^{(\alpha)}$, e.g., the \textit{hypergraph embedding model} (HEM)~\cite{zhen2023_HEM} assumes that $C^{(\alpha)}$ is diagonal, reducing the computational cost to $\mathcal{O}(\alpha d)$.

An alternative way of translating node positions into hyperedge probabilities makes use of the pairwise distances between nodes. Hyperedges can then be sampled independently, conditioned on node coordinates and other parameters. While this approach can be employed in Euclidean space~\cite{gong2023_hypergraphspectralembedding}, the results for graphs suggest that $d$-dimensional hyperbolic latent spaces $\mathbb{H}^d$ can capture heterogeneous node degrees in addition to node similarity~\cite{boguna2021_networkgeometry, krioukov2010_hyperbolicgeometry}. Motivated by these results, Fritz, Yuan, and Schweinberger~\cite{fritz2025_hyperbolic} introduce a model of non-uniform, simple hypergraphs, in which each node $i \in \mathcal{V}$ has a position $x_i \in \mathbb{H}^d$, and the probability $p^{(\alpha)}_{i_1, \dots, i_\alpha}$ of a size-$\alpha$ hyperedge $e$ among nodes $i_1 < i_2 <\dots <i_\alpha$ is
\begin{equation}
    p_{i_1,\dots, i_\alpha}^{(\alpha)}
    =
    \frac{
        2 \eta_\alpha
    }{
        1 + \exp \left ( \Delta_q(\{i_1, \dots, i_\alpha\})\right)
    },
\end{equation}
where $\eta_\alpha \in (0, 1]$ controls how sparse the hypergraph is at hyperedge size $\alpha$, and
\begin{equation}
    \Delta_q(e)
    =
    \left(
        \frac{1}{|e|}
        \sum_{i \in e}
        \left(
            \sum_{j \in e \setminus \{i\}}
            \mathrm{dist}_{d}(x_i, x_j)
        \right)^{q}
    \right)^{1/q}
\end{equation}
summarizes the pairwise distances $\mathrm{dist}_d(x_i,x_j)$ in hyperbolic space through a generalized mean with parameter $q \in \mathbb{R}\setminus\{0\}$. As hyperedges are more probable among nodes that are close in the latent space, this model is suitable for hypergraphs in which interactions are governed by similarity rather than dissimilarity or complementarity. The authors discuss inferring node coordinates and generating hypergraphs. For the latter task, they devise an efficient algorithm that avoids explicitly evaluating all $\binom{N}{\alpha}$ hyperedge probabilities at hyperedge size $\alpha$.

Starting from the bipartite representation of a hypergraph, one can assign both nodes and hyperedges latent positions and model entries of the incidence matrix as independent random variables (conditioned on coordinates), an approach used by Wu, Xu, and Zhu~\cite{wu2025_latentembeddingbipartite}: Each node $i \in \mathcal{V}$ is assigned a latent position $x_i \in \mathbb{R}^d$, and each hyperedge $e \in \mathcal{E}$ a latent position $y_e \in \mathbb{R}^d$, i.e., nodes and hyperedges share the same latent space, with their numbers fixed a priori to $N$ and $E$, respectively. Nodes join hyperedges independently at random with probability,
\begin{equation}
    p_{i,e}
    =
    \frac{
        1
    }{
    1 + \exp
        \left(
             - (\theta_i + \omega_{N,E} + x_i^\mathsf{T} y_e)
        \right)
    }
    ,
\end{equation}
where $\theta_i \in \mathbb{R}$ influences the degree of node $i$, $\omega_{N,E}$ governs the sparsity of the hypergraph's incidence matrix, and nodes are more likely to be members of hyperedges if their latent positions are aligned. In this model of non-uniform hypergraphs, hyperedges are simple but can be parallel. In terms of hypergraph generation, working with the $N \times E$-dimensional incidence matrix can be more efficient than with a high dimensional adjacency tensor. However, the model's independence assumptions are very different from others in this section: Elements of the (weighted) adjacency tensor are no longer independent random variables conditioned on latent variables, but elements of the hypergraph's incidence matrix are.

\subsection{Algorithmically Specified: Growing \& Mechanistic Models}

Many models for hypergraphs with specific mixing patterns are specified in terms of algorithms, e.g., iteratively applied growth rules. This leads to hypergraph models in which hyperedges are often not independent and in which the probability of a hypergraph is, in many cases, not available in closed form. They are nevertheless of great practical relevance as they are often designed for fast hypergraph generation. We focus on models with explicitly specified node attributes. We thus exclude models in which locally more densely connected regions might emerge without explicit assignments of nodes to classes (see, e.g.,~\cite{ko2022_growthpatterns, he2025_hyperedgecopying}), as well as deep generative models, that could potentially learn meso-scale patterns from data (see, e.g.,~\cite{gailhard2025_hygene}).

Here, we introduce a selection of such hypergraph models, highlighting (if applicable) connections to their counterparts for pairwise networks.

To study the interplay of homophily and degree heterogeneity in pairwise networks, several models combine growth by degree-based preferential attachment with class-preferences~\cite{karimi2018_homophilyranking, avin2015_glassceiling, avin2020_mixedpreferentialattachment, dealmeida2013_scalefreehomophilic}. In the case of hypergraphs, the \textit{preferential attachment hypergraphs with high modularity}~\cite{giroire2022_PAhypergraphs} implement this idea: Hypergraphs sampled from this model have heavy-tailed degree distributions and can exhibit different class-mixing patterns. The model permits any number of classes, $C \geq 1$, and its sample space are non-uniform hypergraphs with possibly non-simple and parallel hyperedges. Controlling the hyperedge size distribution and size-dependence of the class-mixing pattern can be challenging, as they are controlled by the interplay of several model parameters.

A model that allows more fine-grained control over the class mixing pattern as well as arbitrary degree distributions is the \textit{hypergraphs with hyperedge homophily} (H3)~\cite{laber2025_effectshigherorderinteractions}. In the H3 model, the number of $(c,\alpha, t)$-hyperedges with respect to an arbitrary reference class $c$ is an input parameter. This fixes the hypergraph homophily in the sense of VBK~\cite{veldt2023_hypergraphhomophily} exactly. Additionally, the expected degree sequence of the model can be specified. The non-uniform hypergraphs sampled from the model are simple. A limitation of the model is that it only supports $C= 2$ classes.

Edge copying is an influential principle in models of pairwise network~\cite{kleinberg1999_webgraph}, and has also been applied to hypergraphs~\cite{he2025_edgecorrelation}. This mechanism can be combined with node labels (i.e., classes) yielding \textit{Copying Hyperedges Influenced by Label Interaction} (CHILI)~\cite{ross2026_growinghypergraphs}, a model in which new hyperedges are noisy copies of existing hyperedges and in which the probability of a node to be included in the new hyperedge depends on its binary class assignment. Hypergraphs from this model can contain parallel hyperedges but nodes do not appear more than once in a given hyperedge, and their degree distribution follows a power law. The authors show how stochastic expectation maximization can be used to infer model parameters.

The \textit{stochastic block hypergraph model}~\cite{pister2024_stochasticblockhypergraph} grows a hypergraph starting from a set of initial hyperedges and the decisions of $N$ nodes to join each of these hyperedges based on a (potentially non-linear) summary of their pairwise affinity to the current hyperedge members. The model is thus not a hypergraph stochastic block model in the sense of Section~\ref{sec:hypergraph-sbm}. While the model supports an arbitrary number of classes, $C \geq 1$, the predetermined binomial degree and hyperedge size distribution limit its flexibility. Parallel hyperedges are possible under the model, but nodes can appear at most once in a hyperedge.

The \textit{hypergraph artificial benchmark for community detection} (h-ABCD)~\cite{kaminski2023_hABCD} creates non-uniform hypergraphs, in which class sizes follow a power-law distribution with user-specified parameters. Note that the number of classes is implicitly determined through the (random) class sizes and the user-specified number of nodes. Hypergraphs from this model have power-law degree distributions that can be tuned by the user. The model distinguishes community hyperedges and background hyperedges. The former carry the hypergraph's class-mixing pattern, while the latter are treated as noise from a community detection perspective. The homogeneity of hyperedges is controlled at each hyperedge size $\alpha$ independently, as the fraction of size-$\alpha$ hyperedges that have $t$ nodes from the class that forms the majority within the hyperedge is an input parameter. Specifying mixing patterns in this way makes the h-ABCD model blind to the exact class labels of nodes constituting a minority within the hyperedge. Two versions of the model, one for simple and one for non-simple hypergraphs, exist. As its name suggests and similar to its graph counterpart~\cite{kaminski2021_ABCD}, the model is designed for fast sampling of benchmark hypergraphs for community detection. However, that the number of classes is random makes it less suitable as a generative model for agent-based simulations.

The \textit{node attribute-based hypergraph generator} (NoAH)~\cite{chun2025_NoAH} generalizes the multiplicative attribute graph model (MAG)~\cite{kim2012_MAG} and differs from the above models, in that nodes are not assigned to a single class but can possess several binary attributes. The pairwise affinities between these attributes determine which hyperedges are created in an iterative procedure. NoAH is designed to leverage node attributes to reproduce features of hypergraph datasets like core-periphery structure and heavy-tailed degree distributions. The associated algorithm NoAHFit can be used to set the model parameters to achieve this goal. Hypergraphs generated with NoAH contain only simple but potentially parallel hyperedges.

Algorithmic hypergraph generators are, of course, not limited to discrete attributes: The \textit{spatial hypergraph model}~\cite{eldaghar2025_spatialhypergraphmodel} assigns each node a position in a Euclidean latent space and groups its $\tilde{d}(i)$ nearest neighbors into hyperedges using the DBSCAN clustering algorithm~\cite{ester1996_DBSCAN}. The model incorporates spatial proximity and, via $\{\tilde{d}(i)\}_{i\in\mathcal{V}}$, a tunable number of direct unique neighbors for each node. Hyperedges in hypergraphs generated using this model can be parallel, but are always simple.

The hypergraph model of Kov\'acs, Benedek, and Palla~\cite{kovacs2025_communitydetectionhypergraphs} relies on a hyperbolic---instead of a Euclidean---latent space and creates size-$\alpha$ hyperedges by connecting a randomly chosen node to its $\alpha - 1$ nearest neighbors in hyperbolic distance. This construction offers control over the number of nodes $N$ and number of hyperedges $E$, as well as over the distribution of node degrees and hyperedge sizes. Hyperedges are naturally simple, and hyperedge rejection ensures that the generated hypergraph does not contain parallel hyperedges. The authors discuss the possibility of using this model to create hypergraphs with locally more densely connected communities through inhomogeneous distribution of nodes in latent space.

Barthelemy proposes an entire family of hypergraph models~\cite{barthelemy2022_randomhypergraphs}, in which hypergraphs are iteratively grown from a set of initial hyperedges. In one instance of the model, dubbed \textit{random spatial hypergraph}, nodes are assigned positions within a disc of fixed radius, and the probability for a node to join a hyperedge depends on a (potentially nonlinear) summary of the pairwise distances between the node and the current members of the hyperedge. This growth process ensures that each node occurs at most once in a given hyperedge; however, the same hyperedge can be created multiple times.

Finally, the group attractiveness model (GAM)~\cite{gallo2024_facetoface} is designed as a model of time-evolving face-to-face interactions based on node positions and node attractiveness. In contrast to other models, node positions change over time in GAM.


\section{Discussion}

In this chapter, we approached higher-order mixing patterns from two angles: Measures of higher-order homophily and hypergraph models for studying mixing patterns. We first surveyed existing measures of higher-order homophily, illustrating each measure through examples and pointing out its prospects and weaknesses. Two key conceptual insights are the absence of a within-/between-class dichotomy for hyperedges and the emergence of new combinatorial constraints on the mixing patterns that could possibly be realized as a hypergraph.

As several higher-order homophily measures have been proposed, researchers are faced with the question of which one to use. A first important consideration is whether node attributes are discrete or continuous. For discrete attributes, measures such as hypergraph homophily or the perplexity-homophily index are suitable, while Chodrow's hypergraph assortativity is an apt choice for continuous node attributes. Another important distinction among measures is the scale of measurement. Some measures quantify mixing patterns on the hypergraph scale (e.g., modularity), others class-wise (e.g., hypergraph homophily), and still others at the level of individual nodes or hyperedges (e.g., MP homophily). Finally, the form in which data is available can restrict the choice of measure. For example, hypergraph homophily can be computed from summary statistics, while RWHS requires access to the exact connectivity pattern of the hypergraph in a data structure that permits simulating random walks and randomizing the hypergraph's structure. These distinctions offer a starting point for selecting measures, but making an informed choice requires reviewing the assumptions and limitations of candidate measures.

Despite the considerable number of different measures in the literature, the study of higher-order mixing patterns remains in its infancy compared to the large body of work on quantifying homophily in networks. This is especially true when it comes to understanding the robustness of measures. Intensive scrutiny has revealed important caveats of classic homophily measures for networks, e.g., in the presence of class imbalance~\cite{karimi2023_inadequacyassortativity}, degree heterogeneity~\cite{cinelli2020_networkconstraints}, or competing link formation mechanisms~\cite{peixoto2022_disentanglinghomophily, sajjadi2024_unveilinghomophily, bachmann2026_PATCH}. Not only are similar limitations likely to apply for hypergraphs, but it is conceivable that the additional structural degrees of freedom in hypergraphs, e.g., nestedness of hyperedges~\cite{landry2024_simpliciality, larock2023_encapsulationstructure}, make the assessment of mixing patterns even more subtle. Reliably applying hypergraph homophily measures means developing a thorough understanding of their limitations.

Turning from measures to models, we provided an introduction to different classes of hypergraph models that are relevant to the study of higher-order mixing patterns. All of these models face the challenge of specifying a distribution on the space of hypergraphs (with node attributes)---complicated combinatorial objects---using few interpretable input parameters or constraints. We showed that different models and their associated sampling algorithms solve this problem in unique ways depending on their intended purpose, emphasizing computational efficiency, realism, and analytical tractability to different degrees. 

For hypergraph stochastic block models and many geometric hypergraph models, the probability of a hypergraph under the model is available in closed form, making them suitable for statistical inference and theoretical studies. However, the large number of possible hyperedges makes na\"ively generating hypergraphs from these models challenging. Hypergraph models specified in terms of algorithmic rules often permit fast generation of hypergraphs with a specific mixing pattern. However, the probability of a hypergraph under these models is typically not available in closed form. As they deviate from the maximum-entropy principle, the sampled hypergraphs are also not maximally random with respect to properties other than the specified mixing pattern; this needs to be kept in mind when using them in simulation studies.
 
Hypergraph models for higher-order mixing patterns ultimately aim to describe ``higher-order structure'', i.e., their goal is to go beyond what could easily be encoded in pairwise edges. Some hypergraph models fall short of this goal by employing effectively pairwise constructions, e.g., growth rules or hyperedge probabilities that are based on repeated pairwise comparisons. These approaches may be analytically convenient and computationally efficient, but they restrict the higher-order mixing patterns a model can express. They might, however, be an interesting starting point for investigating to what extent higher-order interactions are (statistically) reducible to pairwise descriptions. 

In pairwise networks, several phenomena have been discovered that go beyond mixing patterns in a narrow sense but are intimately related to them: One example is \textit{monophily}, an excess variance in the class preference of individuals, which may induce similarity among the neighbors of nodes~\cite{altenburger2018_monophily}. Other examples include the majority illusion~\cite{lee2019_perceptionbias,lerman2016_majorityillusion}, the homophily paradox~\cite{evtushenko2021_paradoxsecondorderhomophily}, and the challenge of distinguishing latent homophily and social influence~\cite{shalizi2011_homophilycontagion}. Investigating such phenomena in the context of higher-order interactions is an important next step in understanding the impact of higher-order mixing patterns on social systems.

As the design of measures and models cannot be an end in itself, higher-order homophily research will eventually have to address the central question: Do higher-order mixing patterns shape social systems (e.g., individual outcomes or the dynamics of social ties) beyond what a pairwise perspective can reveal? While recent modeling studies suggest such effects, e.g., in the context of contagion phenomena~\cite{laber2025_effectshigherorderinteractions, landry2023_opiniondisparity}, gathering empirical evidence in real-world systems is the true frontier of higher-order homophily research. 

\begin{acknowledgement}
We thank Samantha Dies and Phil Chodrow for their valuable comments on the first draft of this chapter. We also thank Nicholas Landry, Juan G. Restrepo, and Fabian Tschofenig for insightful discussions. M.L. was supported by the Inaugural Joseph E.~Aoun Endowment.
\end{acknowledgement}

\bibliographystyle{spphys}
\bibliography{author/bib}

\end{document}